\documentclass[journal,letterpaper]{IEEEtran}
\usepackage{mathtools}
\usepackage{textcomp}
\usepackage{stfloats}
\usepackage{caption}
\usepackage{subcaption}
\usepackage{url}
\usepackage{verbatim}
\usepackage{stackengine}
\usepackage{amsthm}
\usepackage{cite}
\usepackage{scalerel}
\usepackage{longtable,tabularx}
\usepackage{xspace}
\usepackage{color}
\usepackage{bm}
\usepackage{mdframed}
\usepackage{multirow}
\usepackage{float}
\usepackage{array, makecell, booktabs, colortbl}
\usepackage{amsmath, amssymb, amsfonts}
\usepackage{algpseudocode}
\usepackage{siunitx}
\usepackage{enumitem}
\usepackage{xurl}
\usepackage{balance}
\usepackage[most]{tcolorbox}
\usepackage{xcolor}
\usepackage{graphicx}
\usepackage{algorithm}
\newcommand{\ra}[1]{\renewcommand{\arraystretch}{#1}}
\DeclareMathOperator{\SampleMean}{\texttt{SampleMean}}
\DeclareMathOperator{\SampleCov}{\texttt{SampleCov}}

\newcommand{\algstage}[1]{%
    \Statex \hfill
    \textcolor{gray}{\itshape --- #1 ---}
    \hfill\null
}

\newtheorem{rem}{Remark}

\theoremstyle{definition}

\let\originalleft\left
\let\originalright\right
\renewcommand{\left}{\mathopen{}\mathclose\bgroup\originalleft}
\renewcommand{\right}{\aftergroup\egroup\originalright}
\def\p(#1|#2){p\left(#1\,|\,#2\right)}

\title{System Identification for Nonlinear Stochastic State-Space Models Using Ensemble Kalman Inversion with Application to Lithium-Ion Batteries}

\IEEEoverridecommandlockouts
\author{
    Farzaneh Barat, Rachel Carter, Sara Wilson, Huijeong Kim, Xinfan Lin, and Huazhen Fang%
    \thanks{F.~Barat, R.~Carter, and S.~Wilson are with the Department of Mechanical Engineering,
    University of Kansas, Lawrence, KS 66045, USA
    (e-mail: barat@ku.edu; rachel.carter@ku.edu; sewilson@ku.edu).}%
    \thanks{H.~Kim is with the Department of Civil, Environmental \& Architectural Engineering,
    University of Kansas, Lawrence, KS 66045, USA
    (e-mail: hjkim@ku.edu).}%
    \thanks{X.~Lin is with the Department of Mechanical and Aerospace Engineering,
    University of California, Davis, CA 95616, USA
    (e-mail: lxflin@ucdavis.edu).}%
    \thanks{H.~Fang is with the Department of Mechanical Engineering,
    Michigan State University, East Lansing, MI 48824, USA
    (e-mail: hfang@msu.edu).}%
}
\begin{document}
\bstctlcite{IEEEexample:BSTcontrol}
\maketitle

\begin{abstract}
Nonlinear state-space system identification has attracted decades of interest, yet remains challenging because of the interplay among unknown parameters, nonlinear dynamics, and latent state variables. Its difficulty increases dramatically when the state and parameter dimensions increase. To address this challenge, we develop an ensemble Kalman inversion (EnKI)-based approach. This approach leverages EnKI to perform maximum a posteriori parameter estimation through successive local Gaussian approximations, enabling an iterative and incremental search for unknown parameters. The search combines Monte Carlo sampling with Kalman-type updates to evolve an ensemble of parameter samples, thereby offering empirical stability and the ability to handle strongly nonlinear models. The EnKI search is coupled with state estimation conditioned on the latest parameter estimates. This enables parameter identification along with latent state recovery. To improve scalability for large datasets, we further develop a cascaded EnKI (CasEnKI) method, which assimilates subdatasets sequentially to accelerate computation. We validate the proposed methods using a lithium-ion battery model with ten state variables and 18 unknown parameters. Simulation results show that the EnKI approach offers high identification accuracy and empirical stability in parameter search. Further, experimental results illustrate the computational efficiency and scalability of the CasEnKI method for large battery datasets.
\end{abstract}


\section{Introduction}

State-space models (SSMs) offer a powerful way to represent dynamic systems, by unifying a system’s internal state, external input, and measurable output within a clean and coherent mathematical structure. They have found wide application in nearly all engineering fields to enable system simulation, state estimation, and control, since first introduced by Rudolf Kalman in his seminal work~\cite{Kalman:IFAC:1960}. An SSM must be identified from data before being used. While the goal of identification is to determine the model's unknown parameters, it also requires directly or indirectly inferring the typically large number of latent state variables. The resulting large-scale search over the unknown parameters and states often faces significant challenges, including divergence, expensive computation, and poor identifiability. Owing to both its importance and challenges, the identification problem for SSMs has attracted perennial attention over the past decades, but still remains far from fully resolved.  

\subsection{Literature Review on SSM Identification}
Interest in SSM identification dates back to the 1960s. Early studies approached it by including the unknown parameters as extra state variables and applying the extended Kalman filter to estimate both the states and parameters~\cite{Ljung:1977:IEEE,Cox:1964}. This idea of combined state-parameter estimation has since been implemented using different filtering techniques and for different applications~\cite{Ljung:1979:TAC, Ljung:1999, Merwe:2001}. A related approach is the so-called dual estimation, in which two Kalman filters separately estimate the parameters and the states while exchanging information with each other~\cite{Wan:1999, Wan:2000}. However, these methods often exhibit brittle convergence, as the state and parameter estimation errors can mutually amplify over time, and they are also susceptible to weak identifiability and model nonlinearity.

Alternatively, probabilistic estimation methods have become increasingly popular for identifying SSMs. Among them, maximum likelihood (ML) estimation seeks to maximize the likelihood of the observed measurement data conditioned on the unknown parameters, whereas maximum a posteriori (MAP) estimation maximizes the posterior distribution of the parameters given the data. Although formulating ML- or MAP-based parameter estimation problems is conceptually straightforward, solving them is challenging due to the presence of latent state variables. To address this difficulty, the literature has explored several methodological directions as reviewed below. For clarity, the review focuses on ML formulations, as MAP estimation can be viewed as ML augmented with a known prior.

\begin{itemize}
\item {\em Numerical optimization.}
ML estimation leads naturally to optimization problems due to its pursuit of likelihood maximization. For linear Gaussian SSMs, ML cost functions can be computed through the use of Kalman filtering and optimized via gradient-based methods~\cite{SCHON:2011:A, Olsson:2008:B}. For nonlinear SSMs, closed-form ML cost functions are generally unavailable; instead, particle-filter-based likelihood approximations can be formulated, followed by gradient-based parameter search~\cite{doucet:2001:sequential}. Computing these gradients is often difficult, thus motivating the use of stochastic gradient approximations~\cite{Robbins:1951:MS, doucet:2003:parameter}.
  
\item {\em Expectation--maximization (EM) algorithm.}
The EM algorithm performs ML estimation in the presence of latent variables by iteratively alternating between computing the expected complete-data log-likelihood under the current parameter estimates (E-step) and maximizing this expectation to update the parameters (M-step)~\cite{Dempster:1977:RSS}. When applied to identify an SSM, the E-step requires solving a state smoothing problem, and the M-step optimizes over parameters. For linear Gaussian SSMs, EM yields closed-form iterative parameter updates~\cite{Shumway:1982:TSA, Gibson:2005:IEEE}. For nonlinear SSMs, the E-step requires particle smoothers~\cite{SCHON:2011:A, WILLS:2008:IFAC} or sigma-point smoothers~\cite{vaananen:2012, Kokkala:2014:EM}, significantly increasing computational costs.  

\item {\em Markov chain Monte Carlo (MCMC) sampling.}
MCMC methods construct a Markov chain whose stationary distribution matches the target distribution, and then draw samples by running the chain until convergence. Linear SSMs can be identified using the Metropolis--Hastings algorithm, a basic yet popular MCMC method~\cite{hastings:1970:monte}. For nonlinear SSMs, a central difficulty is designing effective proposal distributions. The work in~\cite{andrieu:2010:particle} addresses this by developing particle MCMC, which leverages particle filtering to form reliable proposals. Other advanced MCMC techniques, such as Hamiltonian Monte Carlo~\cite{neal:2011:mcmc} and adaptive Metropolis schemes~\cite{haario:2001:adaptive}, can improve sampling from complex distributions to enhance the identification of nonlinear SSMs.
\end{itemize}

Despite the progress surveyed above, existing methods remain largely confined to low-dimensional nonlinear SSMs with only a handful of parameters; in practice, they are often vulnerable to divergence, local optima, and substantial computational burdens. These limitations stem from the unavoidable use of various approximations, the strong nonconvexity of the associated optimization problems, and the enormous complexity in probability structures arising out of nonlinear SSMs---all ultimately traceable to the aforementioned challenge posed by latent state variables. Given the status quo, we aim to push further toward identifying nonlinear SSMs of order greater than ten and with more than ten unknown parameters.

\subsection{Battery Model Identification}

Although the problem of SSM identification arises across numerous domains, we have particular interest in its application to lithium-ion batteries (LiBs). As one of the most important energy-storage technologies available today, LiBs have become a key enabler of electrified transportation, renewable energy integration, and resilient power grids~\cite{tarascon:2001, SCROSATI:2010}. To ensure their performance and safe operation, model-based battery management systems and algorithms are essential; these, in turn, rely on the ability to extract accurate models from data~\cite{rahn:2013:battery}. In this work, we focus on equivalent circuit models (ECMs), which represent LiBs using circuit analogs comprising resistors, capacitors, and voltage sources and offer desirable computational efficiency for onboard computation. The literature includes two main approaches for ECM identification as reviewed below.

\begin{itemize}
\item {\em Experiment-based model calibration.} 
This approach designs and uses charging and discharging experiments to approximately determine model parameters. For example, trickle-current tests can be used to identify the relationship between the state-of-charge (SoC) and the open-circuit voltage (OCV) of a LiB cell, while pulse tests allow calibration of resistance/capacitance (RC) parameters by fitting the transient responses~\cite{Chen:2006:IEEE, Zhang:2020:en}. This approach is effective for identifying simple models such as the Rint model or the Thevenin model with one or two RC pairs. To characterize temperature dependence, one can conduct tests under different thermal conditions to quantify how internal resistance varies with temperature~\cite{LUDWIG:2021:PS}. For more sophisticated models, it is possible to design multi-pronged experiments to excite different dynamic processes within a LiB cell, so as to determine the corresponding parameters~\cite{BIJU:2023:AE}. In general, experiment-based calibration is straightforward to understand and implement. However, its accuracy is often limited, and parameter estimation errors may accumulate when multiple tests are used to calibrate different parameters within a model.

\item {\em Optimization-based model identification.} Another approach lies in data fitting, which seeks to minimize the discrepancy between model predictions and measurement data. This approach often leads to least-squares formulations~\cite{FENG:2015:PS, Sitterly:2011:IEEE, TIAN:2020:ES}. Such problems can be solved easily when the excitation input is a constant current, since the model’s response admits an explicit parameterization~\cite{TIAN:2020:ES}. However, constant-current excitations provide limited amounts of information about the unknown parameters and thus constrain estimation accuracy. Richer excitation signals, like variable charging/discharging profiles, generate more informative data to help improve parameter estimation~\cite{Lin:2017:IEEE, SONG:2018:AE}. To leverage such data, the study in~\cite{Tian:2021:TCS} exploits the prediction-error method for a LiB model with linear state dynamics. For nonlinear ECMs, the resulting optimization problems are nonlinear, nonconvex, and high-dimensional due to latent state variables. Gradient-based optimization methods often struggle with these challenges. Hence, derivative-free methods including evolutionary algorithms, such as genetic algorithms~\cite{MALIK:2014:PS}, particle swarm optimization~\cite{YU:2017:IEEE}, and differential evolution~\cite{YANG:2014:PS}, and Bayesian optimization~\cite{TU:2026:arXiv} have proven useful. These methods, however, are computationally intensive, and their practical performance depends on initialization, population size, and iteration budgets, among others.
\end{itemize}

ECMs present themselves in the form of SSMs, so a natural way for identifying them is through SSM identification. This perspective enables the development of identification methods that are principled in design, applicable to arbitrary charging/discharging profiles, computationally tractable, and capable of explicitly accounting for both process and measurement noises in a model. Despite these advantages, few studies have adopted this perspective for ECMs, leaving a gap that this work seeks to address.

\subsection{Statement of Contributions}

In this paper, we consider MAP-based parameter estimation for nonlinear stochastic SSMs. To address this problem, we exploit ensemble Kalman inversion (EnKI). EnKI is an ensemble-based parameter estimation technique. Specifically, it starts from an ensemble drawn from a prior parameter distribution, and then propagates the ensemble via iterative Kalman-style corrections to progressively approximate the posterior parameter distribution~\cite{Iglesias:2013:IP}. This technique, by design, combines Monte Carlo sampling with the Kalman-type update to enable gradient-free, empirically stable, and computationally manageable parameter search to tackle even complex problems~\cite{Kovachki:2019:IP}. Tapping into this potential, this paper makes the following contributions:

\begin{itemize}

\item We develop an EnKI-based approach for nonlinear SSM identification. The proposed approach intertwines EnKI for parameter estimation with the ensemble Kalman filter (EnKF) for latent state estimation, alternating between the two until convergence.

\item We improve the computational scalability of the proposed approach for large datasets by partitioning the data into smaller subsets and applying EnKI sequentially across them. This extension, termed cascaded EnKI (CasEnKI), considerably reduces the computational cost, while enabling the use of large datasets to ensure accurate identification of high-order SSMs with many unknown parameters.

\item We apply the proposed CasEnKI approach to a LiB model called BattX, which is tenth-order and has 18 unknown parameters. The simulation and experimental results not only validate the effectiveness of this approach, but also show its power for addressing practically challenging applications in battery engineering and beyond.

\end{itemize}

A preliminary version of this work appeared in~\cite{Barat:2026}. This paper substantially extends that earlier study in several aspects, including: 1) the consideration of stochastic SSMs, rather than only deterministic SSMs, together with the corresponding modifications to the identification approach; 2) the development of the CasEnKI identification approach to handle large datasets; and 3) the validation on a higher-order LiB model with many more parameters.

\subsection{Organization}
The remainder of this paper is organized as follows. 
Section~\ref{sec:EnKI} presents the nonlinear stochastic SSM identification problem and the EnKI-based solution. Section~\ref{sec:CasEnKI} develops CasEnKI for scalable identification from large datasets. Section~\ref{sec:BattX} validates the CasEnKI approach on identification of the BattX model through both simulations and experiments. Finally, Section~\ref{sec:con} concludes the paper, and a few useful methods and results are summarized in the appendices.

\section{EnKI-Based SSM Identification}
\label{sec:EnKI}

In this section, we begin by introducing the system identification problem for SSMs and then proceed to develop the EnKI-based solution.

\subsection{Problem Formulation}
Consider a discrete-time stochastic nonlinear SSM of the form:
\begin{subequations}
\label{eq:ssm_model}
\begin{align}
\bm{x}_{k+1}
&=
f(\bm{x}_k,\bm{u}_k,\bm{\theta})+\bm{w}_k,
\label{eq:ssm_state}\\
\bm{y}_k
&=
h(\bm{x}_k,\bm{u}_k,\bm{\theta})+\bm{v}_k,
\label{eq:ssm_obs}
\end{align}
\end{subequations}
where $\bm{x}_k \in \mathbb{R}^{n_x}$ denotes the state, $\bm{u}_k \in \mathbb{R}^{n_u}$ is the input, $\bm{y}_k \in \mathbb{R}^{n_y}$ is the measured output, and $\bm{\theta} \in \mathbb{R}^{n_\theta}$ is the unknown parameter vector to be identified. 
Further, $\bm{w}_k$ and $\bm{v}_k$ are the mutually independent process and measurement noises. Both are white Gaussian with $\bm{w}_k {\sim} \mathcal{N}(\bm{0},\bm{Q})$ and $\bm{v}_k {\sim} \mathcal{N}(\bm{0},\bm{R})$, where $\bm{Q}$ and $\bm{R}$ are the respective covariance matrices. In the sequel, we omit $\bm{u}_k$ whenever no confusion arises, for notational simplicity.

Given~\eqref{eq:ssm_model}, the measurement dataset is
\begin{align*}
\bm Y
=
\begin{bmatrix}
\bm y_1^\top &
\bm y_2^\top &
\cdots &
\bm y_K^\top
\end{bmatrix}^{\top},
\end{align*}
where $K$ is the terminal time. We also use the shorthand notation $\bm y_{1:k} = [\bm y_1^\top,\ldots,\bm y_k^\top]^\top$; in particular, $\bm Y = \bm y_{1:K}$.

The objective of system identification is to estimate the unknown $\bm\theta$ from $\bm Y$. To this end, we adopt an MAP framework by considering
\begin{align}
\widehat{\bm{\theta}}_{\mathrm{MAP}} = \arg\max_{\bm{\theta}} \p(\bm{\theta}|\bm{Y}),
\label{eq:MAP}
\end{align}
where $\p(\bm{\theta}|\bm{Y})$ is  the posterior distribution of $\bm\theta$ conditioned on $\bm{Y}$. As~\eqref{eq:MAP} indicates, the optimal estimate $\widehat{\bm{\theta}}_{\mathrm{MAP}}$ is the most probable value of $\bm\theta$ given $\bm{Y}$. By Bayes' rule,
\begin{align*}
\p(\bm{\theta}|\bm{Y}) 
\propto 
\p(\bm{Y}|\bm{\theta}) p(\bm{\theta}),
\end{align*}
where $\p(\bm{Y}|\bm{\theta})$ is the likelihood of $\bm Y$, and $p(\bm\theta)$ is the prior of $\bm\theta$. This implies that $\widehat{\bm{\theta}}_{\mathrm{MAP}}$ naturally strikes a balance between fitting the observed data through the likelihood term and incorporating prior knowledge through the prior term.

The above SSM identification problem is challenging for two key reasons. First, $\p(\bm{\theta}|\bm{Y})$ generally admits no closed-form expression. Second, the state trajectory $\bm x_{0:K}$ is latent, in addition to the
unknown $\bm\theta$, making it hard to evaluate
$\p(\bm\theta|\bm Y)$. To overcome these challenges, we develop an EnKI-based approach next.

\subsection{EnKI-Based System Identification}

EnKI performs ensemble-based MAP estimation of unknown parameters, inspired by the well-known ensemble Kalman filtering method~\cite{Kovachki:2019:IP,Iglesias:2013:IP}. We can interpret this technique as iteratively refining a local Gaussian approximation for the posterior parameter distribution $\p(\bm{\theta}|\bm{Y})$. The approximation evolves toward the most probable mode of $\p(\bm{\theta}|\bm{Y})$ as the iterations proceed. A more detailed explanation is as follows.

To approximate $\p(\bm{\theta}|\bm{Y})$, we first introduce a Gaussian approximation of the joint distribution $p(\bm{\theta},\bm{Y})$ by:
\begin{align}
\label{eq:Joint_Gaussian}
    \begin{bmatrix}
        \bm{\theta} \\
        \bm{Y}
    \end{bmatrix}
     \sim  \mathcal{N}
    \left(
    \begin{bmatrix}
        \overline{\bm{\theta}} \\
        \overline{\bm{Y}}
    \end{bmatrix},
    \begin{bmatrix}
        \bm{C}^{\bm{\theta\theta}} & \bm{C}^{\bm{\theta Y}} \\
        (\bm{C}^{\bm{\theta Y}})^\top & \bm{C}^{\bm{YY}} + \bm\Gamma
    \end{bmatrix}
    \right),
\end{align}
where $\overline{\bm{\theta}}$ and $ \overline{\bm{Y}}$ are the means and 
$\bm{C}^{\bm{\theta\theta}}, \bm{C}^{\bm{\theta Y}}, \bm{C}^{\bm{YY}}$ are the empirical covariance and cross-covariance matrices. Here, $\bm \Gamma$ accounts for the measurement noise and the uncertainty due to the latent state trajectory. Hence, $\bm \Gamma \succeq \overline{\bm R}$,  where $\overline{\bm R} = \bm I_K\otimes\bm R$, and it can be made tunable by letting 
$\bm\Gamma
=
\lambda_{\Gamma}\overline{\bm R}
$ with $
\lambda_{\Gamma}\geq 1$ is a tunable covariance-inflation factor. 
From~\eqref{eq:Joint_Gaussian}, a Gaussian approximation $\p(\bm{\theta}|\bm{Y})$ results:
\begin{align*}
\bm{\theta} \, | \, \bm{Y}  \sim  \mathcal{N}(\bm{m}, \bm{P}),
\end{align*}
with
\begin{align*}
\bm m &= \overline{\bm{\theta}} + 
\bm{C}^{\bm{\theta Y}} \big(\bm{C}^{\bm{YY}} + \bm\Gamma\big)^{-1} 
\big(\bm{Y} - \overline{\bm{Y}}\big), \\
\bm P &= \bm{C}^{\bm{\theta\theta}} - 
\bm{C}^{\bm{\theta Y}} \big(\bm{C}^{\bm{YY}} + \bm\Gamma\big)^{-1} 
\big(\bm{C}^{\bm{\theta Y}}\big)^\top.
\end{align*}
Given~\eqref{eq:Joint_Gaussian}, we consider two ensembles of samples, $\{\bm\theta^{i}\}_{i=1}^{N_e}$ and $\{\bm Y^{i}\}_{i=1}^{N_e}$, which together form an empirical joint distribution for $p(\bm{\theta}, \bm Y)$. Here, $N_e$ is the number of samples in the ensembles. Based on these ensembles, we compute the following empirical means and covariances:
\begin{align*}
\overline{\bm{\theta}}
&=
\SampleMean\left(\bm{\theta}^{1\rightarrow N_e}\right),
\\
\overline{\bm{Y}}
&=
\SampleMean\left(\bm{Y}^{1\rightarrow N_e}\right),
\\
\bm{C}^{\bm{\theta\theta}}
&=
\SampleCov\left(\bm{\theta}^{1\rightarrow N_e}\right),
\\
\bm{C}^{\bm{\theta Y}}
&=
\SampleCov\left(
\bm{\theta}^{1\rightarrow N_e},
\bm{Y}^{1\rightarrow N_e}
\right),
\\
\bm{C}^{\bm{YY}}
&=
\SampleCov\left(\bm{Y}^{1\rightarrow N_e}\right),
\end{align*}
where $\SampleMean$ and $\SampleCov$ are defined in Appendix~\ref{app:sample_mean_cov}. 

The above shows an ensemble-based update procedure to identify $\bm \theta$; $\bm m$ represents the estimate of $\bm \theta$ extracted from $\bm Y$, and $\bm P$ is the associated covariance. However, the Gaussian approximation in~\eqref{eq:Joint_Gaussian} is global in nature and has limited fidelity in representing the true distribution $p(\bm \theta, \bm Y)$ due to the model nonlinearity. This will significantly compromise the estimation accuracy for $\bm\theta$. To remedy the issue, we seek to impose successive local Gaussian approximations instead. To this end, we write
\begin{align}
\label{eq:EnKI-full-posterior}
\p(\bm\theta|\bm Y) 
\propto 
\p(\bm Y|\bm\theta)\,p(\bm\theta) =
\left(
\prod_{\ell=0}^{L-1}
\big[\p(\bm Y|\bm\theta)\big]^{\alpha_\ell}
\right)
p(\bm\theta),
\end{align}
where $\ell$ is an integer index, $\alpha_\ell$ is a tempering parameter satisfying $0<\alpha_\ell\leq 1$ and
$\sum_{\ell=0}^{L-1}\alpha_\ell=1$. The factor
$\big[\p(\bm Y|\bm\theta)\big]^{\alpha_\ell}$ represents a tempered likelihood.
We define the sequence of tempered posteriors as
\begin{align}
p_{\ell+1}(\bm{\theta}\,| \,\bm{Y})
\propto
\left[\p(\bm{Y}|\bm{\theta})\right]^{\alpha_\ell}
p_\ell(\bm{\theta}\,| \,\bm{Y}),
\label{eq:tempered_posterior}
\end{align}
where $\ell=0,\ldots,L-1$, 
with $p_0(\bm\theta\,|\,\bm Y)=p(\bm\theta)$. As $\ell$ increases, $p_\ell(\bm\theta\,|\,\bm Y)$ is progressively updated, moving closer to the target posterior $\p(\bm \theta|\bm Y)$. When $\ell=L$, $p_\ell(\bm\theta\,|\,\bm Y)$ coincides with $\p(\bm\theta|\bm Y)$. 
Now, we impose a local Gaussian approximation  around $\overline{\bm \theta}_\ell$:
\begin{align*}
\begin{bmatrix}
\bm\theta\\
\bm Y
\end{bmatrix}
{\sim}
\mathcal N\left(
\begin{bmatrix}
\overline{\bm\theta}_\ell\\
\overline{\bm Y}_\ell
\end{bmatrix},
\begin{bmatrix}
\bm C_\ell^{\bm\theta\bm\theta}
&
\bm C_\ell^{\bm\theta\bm Y}
\\
\left(
\bm C_\ell^{\bm\theta\bm Y}
\right)^\top
&
\bm C_\ell^{\bm Y\bm Y}
+
\alpha_\ell^{-1}\bm\Gamma
\end{bmatrix}
\right),
\end{align*}
where $\alpha_\ell^{-1} \bm\Gamma$ results from the tempered likelihood  $\left[\p(\bm{Y}|\bm{\theta})\right]^{\alpha_\ell}$. Then, $p_{\ell+1}(\bm\theta\,|\,\bm Y)$ admits a local Gaussian approximation: 
\begin{align*}
p_{\ell+1}(\bm\theta\,|\,\bm Y)
\approx
\mathcal N\left(
\overline{\bm\theta}_{\ell+1},
\bm P_{\ell+1}
\right),
\end{align*}
where
\begin{subequations}
\label{eq:EnKI_Gaussian_update}
\begin{align}
\overline{\bm{\theta}}_{\ell+1}
&= \overline{\bm{\theta}}_{\ell}
 + \bm{C}^{\bm{\theta Y}}_{\ell}
 \left( \bm{C}^{\bm{YY}}_{\ell} + \alpha_{\ell}^{-1}\bm\Gamma \right)^{-1}
 \left( \bm{Y} - \overline{\bm{Y}}_{\ell} \right),
\label{eq:EnKI_mean_update_iter} \\
\bm{C}^{\bm\theta\bm\theta}_{\ell+1}
&=
\bm{C}^{\bm{\theta\theta}}_\ell
-
\bm{C}^{\bm{\theta Y}}_\ell
\left(
\bm{C}^{\bm{YY}}_\ell+\alpha_\ell^{-1}\bm\Gamma
\right)^{-1}
\left(\bm{C}^{\bm{\theta Y}}_\ell\right)^\top.
\label{eq:EnKI_cov_update_iter}
\end{align}
\end{subequations}

To implement the procedure in~\eqref{eq:EnKI_Gaussian_update}, we  introduce a parameter ensemble, $\{ \bm \theta_\ell^i \}_{i=1}^{N_e}$, and an output ensemble, $\{\bm Y_\ell^i \}_{i=1}^{N_e}$, to empirically approximate the distributions of $\bm \theta$ and $\bm Y$. By~\eqref{eq:EnKI_Gaussian_update}, $\bm \theta_\ell^i$ for $i=1,\ldots,N_e$ can be updated iteratively as follows:
\begin{align}
\bm\theta_{\ell+1}^{i}
&=
\bm\theta_\ell^i
+
\bm C_\ell^{\bm\theta\bm Y}
\left( \bm C_\ell^{\bm Y\bm Y}
+
\alpha_\ell^{-1}\bm\Gamma
\right)^{-1} \left( \bm Y
-
\bm Y_\ell^i
\right),
\label{eq:EnKI_parameter_update}
\end{align}
where
\begin{subequations}
\label{eq:EnKI_iter}
\begin{align}
\overline{\bm{\theta}}_\ell
&=
\SampleMean\left(\bm{\theta}_\ell^{1\rightarrow N_e}\right),
\\
\overline{\bm{Y}}_\ell
&=
\SampleMean\left(\bm{Y}_\ell^{1\rightarrow N_e}\right),
\label{eq:EnKI_mean_iter}
\\
\bm{C}^{\bm{\theta\theta}}_\ell
&=
\SampleCov\left(\bm{\theta}_\ell^{1\rightarrow N_e}\right),
\label{eq:EnKI_covth_iter}
\\
\bm{C}^{\bm{\theta Y}}_\ell
&=
\SampleCov\left(
\bm{\theta}_\ell^{1\rightarrow N_e},
\bm{Y}_\ell^{1\rightarrow N_e}
\right),
\label{eq:EnKI_crosscov_iter}
\\
\bm{C}^{\bm{YY}}_\ell
&=
\SampleCov\left(\bm{Y}_\ell^{1\rightarrow N_e}\right).
\label{eq:EnKI_covY_iter}
\end{align}
\end{subequations}
After the final iteration, we obtain the final parameter estimate: $\widehat{\bm{\theta}} =
\overline{\bm{\theta}}_{L}$, along with the covariance $\bm C^{\bm\theta\bm\theta} = \bm C^{\bm\theta\bm\theta}_{L}$.

The above procedure constitutes the EnKI method, which iteratively updates the ensemble that approximately represents $\p(\bm \theta|\bm Y)$ until achieving convergence. By leveraging tempered posteriors and local Gaussian approximations, EnKI incrementally refines the posterior estimate. Further, the ensemble representation, together with the Kalman-type update, enables a derivative-free and computationally tractable inference procedure. Some further remarks are as follows.

\begin{rem}
We interpret the EnKI method as resulting from successive local Gaussian approximations for $p\left(\bm \theta, \bm Y\right)$ that yield an iterative procedure. The local approximation overcomes a key issue in the global approximation in~\eqref{eq:Joint_Gaussian}---overconfidence and consequently, strongly biased estimation. This feature substantially reduces---though does not eliminate---biases and helps improve the stability in estimation. 
\end{rem}

\begin{rem}
The literature has suggested that the EnKI method enjoys empirical stability~\cite{DUFFIELD:2022:SPL}. For nonlinear models, if the span of the initial ensemble covers the true value of $\bm \theta$, then the subsequent ensembles during the iteration can maintain the coverage, rather than diverge, even for high-dimensional problems. With this merit, the EnKI method usually converges to meaningful estimates despite certain biases.
\end{rem}

\begin{rem}
The choice of the tempering parameter $\alpha_\ell$   is important for the iteration speed and convergence. In practice, $\alpha_\ell$ is often selected to be small enough to ensure convergence. 
However, there are some alternative ways to choose $\alpha_\ell$ to speed up convergence. An example is a bisection search strategy proposed in~\cite{Kantas:2014:SIAM}. In this paper, we adopt an alternative technique---the data misfit controller proposed in~\cite{Iglesias:2021:IP}. Given $\{\bm Y_\ell^i\}_{i=1}^{N_e}$, this technique considers $\Phi_\ell^{i}$, $\overline{\Phi}_\ell$, and  $\sigma_{\Phi_\ell}^2$, which denote the empirical misfit of each ensemble member, and  the empirical mean and variance of the ensemble misfits, respectively:
\begin{align*}
\Phi_\ell^{i}
&=
\frac{1}{2}
\left(
\bm Y-\bm Y_\ell^i
\right)^\top
\bm\Gamma^{-1}
\left(
\bm Y-\bm Y_\ell^i
\right), \\
\overline{\Phi}_\ell
&=
\frac{1}{N_e}
\sum_{i=1}^{N_e}
\Phi_\ell^{i}, \quad 
\sigma_{\Phi_\ell}^2
 =
\frac{1}{N_e-1}
\sum_{i=1}^{N_e}
\left(
\Phi_\ell^{i}-\overline{\Phi}_\ell
\right)^2.
\end{align*}
Then, $\alpha_\ell$ is selected as
\begin{align}
\alpha_\ell
=
\min\left(
\max\left(
\frac{d_{\bm Y}}{2\overline{\Phi}_\ell},
\sqrt{\frac{d_{\bm Y}}{2\sigma_{\Phi_\ell}^2}}
\right),
1-t_\ell
\right),
\label{eq:DMC_tempering_parameter}
\end{align}
where $d_{\bm Y} =K n_y$ is the dimension of $\bm Y$ and $t_\ell=\sum_{r=0}^{\ell-1}\alpha_r$ is the cumulative tempering parameter.
\end{rem}

\subsection{Construction of the Output Trajectory Ensemble}
\label{subsec:predicted_output_ensemble}
To implement~\eqref{eq:EnKI_iter}, we must construct the output trajectory ensemble
$\{\bm Y_\ell^{i}\}_{i=1}^{N_e}$, which approximates the likelihood
distribution $\p(\bm Y|{\bm\theta}_\ell^{i})$. 
To address this, we consider $\p(\bm Y|\bm\theta)$ and factorize it using Bayes' rule:
\begin{align*}
\p(\bm Y|\bm\theta)=\p(\bm y_{1:K}|\bm\theta) = \prod_{k=1}^{K-1}
\p(\bm y_{k+1}|\bm y_{1:k},\bm\theta) \cdot \p(\bm y_1|\bm\theta). 
\end{align*} 
Consequently, evaluating the complete likelihood $\p(\bm y_{1:K}|\bm\theta)$  reduces to computing the predictive output distribution $\p(\bm y_{k+1}|\bm y_{1:k},\bm\theta)$. For the SSM in~\eqref{eq:ssm_model}, this distribution is given by 
\begin{align}
\label{Predictive-output-distribution}
\p(\bm y_{k+1}|\bm y_{1:k},\bm\theta)
&=
\int
\p(\bm y_{k+1}|\bm x_{k+1},\bm\theta)\nonumber\\
& \qquad\qquad\cdot
\p(\bm x_{k+1}|\bm y_{1:k},\bm\theta)
\,d\bm x_{k+1}.
\end{align}
Here, $\p(\bm x_{k+1}|\bm y_{1:k},\bm\theta)$ is the predictive state distribution, satisfying the recursive Bayesian
filtering equations: 
\begin{subequations}\label{Bayesian-filtering-equations}
\begin{align}
    \p(\bm x_{k+1}|\bm y_{1:k},\bm\theta)
    &=
    \int
    \p(\bm x_{k+1}|\bm x_k,\bm\theta)
    \p(\bm x_k|\bm y_{1:k},\bm\theta)
    \,d\bm x_k,\\
    \p(\bm x_k|\bm y_{1:k},\bm\theta)
    &\propto
    \p(\bm y_k|\bm x_k,\bm\theta)
    \p(\bm x_k|\bm y_{1:k-1},\bm\theta).
\end{align}
\end{subequations}
As shown by~\eqref{Predictive-output-distribution}--\eqref{Bayesian-filtering-equations}, constructing $\{\bm Y_\ell^i\}_{i=1}^{N_e}$  amounts to first performing state filtering conditioned on ${\bm\theta}_\ell^{i}$ and then  using the resulting state estimates to predict the outputs. 
Various Bayesian filtering methods, including nonlinear Kalman filters and 
particle filters, are available for this purpose~\cite{Fang:2018:IEEE}. In this work, we adopt the EnKF method. This method, like EnKI, performs ensemble-based estimation but for the latent states. 
At each time step, it assimilates the new measurement $\bm y_k$ to obtain an updated state ensemble representing   $\p(\bm x_{k}|\bm y_{1:k}, {\bm\theta}_\ell^i )$ and then  propagates the state ensemble to approximate $\p(\bm x_{k+1}|\bm y_{1:k}, {\bm\theta}_\ell^i )$. The resulting state ensemble is subsequently used to construct the output ensemble for $\p(\bm y_{k+1}|\bm y_{1:k}, {\bm\theta}_\ell^i)$. Averaging each output ensemble  for $k=1,\ldots, K$ and collecting them all will then yield $\{\bm Y_\ell^i\}_{i=1}^{N_e}$ needed by EnKI.

The EnKF procedure specifically runs as follows. Given $\bm \theta_\ell^i$, suppose that an ensemble $\{\bm{x}_{k|k-1}^{j,i}\}_{j=1}^{N_s}$ is available to approximate $\p(\bm x_k|\bm y_{1:k-1}, \bm \theta_\ell^i)$, where $N_s$ is the ensemble size. Here, the iteration index $\ell$ is omitted from   $\bm{x}_{k|k-1}^{j,i}$ for notational simplicity.  Using $\bm y_k$,  the EnKF applies the Kalman-type update to this ensemble: 

\begin{subequations}
\label{eq:EnKF_sate_filtering_output_prediction}
\begin{align}
\label{eq:state_ensemble_correction}
\bm{x}_{k|k}^{j,i}
&=
\bm{x}_{k|k-1}^{j,i}
+
\bm{K}_{k}^{i}
\left(
\bm{y}_k
-
\bm{y}_{k|k-1}^{j,i}
\right), \\
\label{eq:EnKF_predicted_measurement}
\bm{y}_{k|k-1}^{j,i}
&=
h\left(
\bm{x}_{k|k-1}^{j,i},\bm{\theta}_\ell^{i}
\right),\\
\bm{K}_{k}^{i}
&=\bm{C}_{k}^{\bm{xy}, i}
\left(
\bm{C}_k^{\bm{yy}, i} +\bm R
\right)^{-1}, \\
\bm{C}_{k}^{\bm{xy}, i}
&=
\SampleCov
\left(
\bm{x}_{k|k-1}^{1\rightarrow N_s,i},
\bm{y}_{k|k-1}^{1\rightarrow N_s,i}
\right),\\
\bm{C}_{k}^{\bm{yy}, i}
&=
\SampleCov
\left(
\bm{y}_{k|k-1}^{1\rightarrow N_s,i}
\right).
\end{align}
\end{subequations}
Here, $\bm K_k^{i}$ is the Kalman gain.
The new ensemble $\{\bm{x}_{k|k}^{j,i}\}_{j=1}^{N_s}$ approximately represents $\p(\bm x_k|\bm y_{1:k}, \bm \theta_\ell^i)$. It then propagates forward through the SSM's dynamics as follows, leading to an ensemble $\{\bm{x}_{k+1|k}^{j,i}\}_{j=1}^{N_s}$ that represents $\p(\bm x_{k+1}|\bm y_{1:k}, { \bm  \theta}_\ell^{i})$:
\begin{align}
\bm{x}_{k+1|k}^{j,i}
=
f\bigl(
\bm{x}_{k|k}^{j,i},\bm{\theta}_\ell^{i}
\bigr)
+
\bm{w}_{k}^{j,i},  
\label{eq:state_ensemble_prediction}
\end{align} 
where $\bm w_k^{j,i}{\sim}\mathcal N(\bm 0,\bm Q)$.

The execution of the EnKF method in~\eqref{eq:EnKF_sate_filtering_output_prediction}--\eqref{eq:state_ensemble_prediction}
produces the output ensemble $\{\bm y_{k|k-1}^{j,i}\}_{j=1}^{N_s}$ (see~\eqref{eq:EnKF_predicted_measurement}), which represents $\p(\bm y_k|\bm y_{1:k-1},\bm\theta_\ell^i)$. Finally, the  output ensembles at each $k$ are averaged and collected to form  
$\bm Y_\ell^i$: 
\begin{align}
\bm Y_\ell^i = \left\{ \frac{1}{N_s} \sum_{j=1}^{N_s} \bm y_{k|k-1}^{j,i}, \ k=1,\ldots,K \right\}.   \label{eq:output_trajectory_average}
\end{align}
The resulting collection $\{\bm Y_\ell^{i}\}_{i=1}^{N_e}$ constitutes the output trajectory ensemble required by the EnKI update in~\eqref{eq:EnKI_iter}.

\begin{rem}
If the SSM in~\eqref{eq:ssm_model} has deterministic dynamics (i.e., $\bm w_k=\bm 0$) and a known initial state $\bm x_0$, the state and output ensembles can be computed directly as below, bypassing the need for Bayesian filtering:
\begin{align*}
\bm{x}_{k+1}^{j,i}
&=
f\bigl(
\bm{x}_{k}^{j,i},\bm{\theta}_\ell^{i}
\bigr),\\
\bm{y}_{k+1}^{j,i}
&=
h\bigl(
\bm{x}_{k+1}^{j,i},\bm{\theta}_\ell^{i}
\bigr)
+
\bm v_{k+1}^{j,i},
\qquad
\bm v_{k+1}^{j,i}{\sim}\mathcal N(\bm 0,\bm R).
\end{align*}
Such a simplification in this specific case substantially lowers the computational cost of constructing $\{\bm Y_\ell^{i}\}_{i=1}^{N_e}$. 
\end{rem}

\subsection{Algorithm Summary and Remarks}

The proposed EnKI-based SSM identification algorithm is summarized in Algorithm~\ref{alg:EnKI}. The algorithm proceeds iteratively. At each iteration $\ell$,   EnKF performs Bayesian state filtering to construct the output trajectory ensembles $\{\bm Y_\ell^i\}_{i=1}^{N_e}$ conditioned on the current parameter ensemble $\{\bm \theta_\ell^i\}_{i=1}^{N_e}$; then, EnKI uses these output ensembles along with the measurement data to update the parameter ensemble. The procedure continues until reaching the maximum iteration number.

Ensembles are capable of approximating complex probability distributions, and the Kalman-type update of ensembles results in a derivative-free estimation procedure that avoids explicit gradient computation. Together, these two features enable the proposed algorithm to effectively identify nonlinear, high-dimensional SSMs while maintaining favorable computational efficiency. Furthermore,   ensemble-based computation is naturally parallelizable, making the algorithm well suited for parallel processing, though this is beyond the scope of this paper.

\begin{algorithm}[htbp]
\caption{EnKI-Based Nonlinear   SSM Identification}
\label{alg:EnKI}
\small
\begin{algorithmic}[1]

\algstage{Initialization}
\State Draw the initial parameter ensemble
$\{\bm\theta_0^i\}_{i=1}^{N_e}{\sim} p(\bm\theta)$.

\vspace{3pt}
\algstage{EnKI-based iterative parameter identification}
\State Set $\ell=0$ and $t_0=0$.
\While{$t_\ell<1$}

    \vspace{3pt}
    \algstage{State filtering }
    \For{$i=1,\ldots,N_e$}
        \State Initialize the state ensemble for each $\bm \theta_\ell^i$
        \State Run the  EnKF method using \eqref{eq:EnKF_sate_filtering_output_prediction}-- \eqref{eq:state_ensemble_prediction} 
        \State Construct $\bm Y_\ell^i$ using \eqref{eq:output_trajectory_average} 
    \EndFor

    \vspace{3pt}
    \algstage{Parameter ensemble update}
    \State Compute
    $\overline{\bm\theta}_\ell$,
    $\overline{\bm Y}_\ell$,
    $\bm C_\ell^{\bm\theta\bm\theta}$,
    $\bm C_\ell^{\bm\theta\bm Y}$, and
    $\bm C_\ell^{\bm Y\bm Y}$
    using \eqref{eq:EnKI_iter} 
    \State Compute 
    $\alpha_\ell$
    using \eqref{eq:DMC_tempering_parameter} 
    \For{$i=1,\ldots,N_e$}
        \State Update $\bm\theta_{\ell+1}^i$
        using \eqref{eq:EnKI_parameter_update} 
    \EndFor
    \State Set $t_{\ell+1}=t_\ell+\alpha_\ell$
    \State Set $\ell\gets\ell+1$
\EndWhile
\State Set $L=\ell$

\vspace{3pt}
\algstage{Final estimate}
\State Set
$\widehat{\bm\theta}
=
\overline{\bm\theta}_{L}$ 
\State \Return $\widehat{\bm\theta}$ 

\end{algorithmic}
\end{algorithm}


\section{CasEnKI for SSM Identification Using Large Datasets}
\label{sec:CasEnKI}

 In Section~\ref{sec:EnKI}, we have developed the EnKI-based algorithm for SSM identification. However, SSM identification often requires large datasets to ensure accuracy, for which the proposed algorithm can become computationally expensive. The main bottleneck lies in computing  $(\bm C_{\ell}^{\bm Y\bm Y}+\alpha_{\ell}^{-1}\bm\Gamma)^{-1}$ as shown in~\eqref{eq:EnKI_parameter_update}---due to the inversion, the computational complexity is $\mathcal O(K^3n_y^3)$ since the measurement dataset $\bm Y\in\mathbb R^{Kn_y}$. This raises an important question: {\em  How can we scale up the proposed algorithm  for large datasets?}

Large datasets generally arise in two scenarios:
\begin{itemize}
\item a single dataset $\bm Y$ containing a long sequence of measurements (i.e., when $K$ is large); and
\item multiple separate datasets.
\end{itemize}
Here, we focus on the first scenario, of which the second can be viewed as a special case. Our solution builds on a divide-and-conquer strategy, which divides $\bm Y$ into segments or subdatasets  and then runs a cascade of incremental EnKI updates over the subdatasets. We refer to  the resulting algorithm as CasEnKI,  describing its  detailed design   below.

To begin with, let us partition $\bm Y$ into $M$ subdatasets, $\{ \bm Y_1, \bm Y_2, \ldots, \bm Y_M\}$. Here,  $\bm Y_m = \left \{ \bm y_k, T_m^0 \leq k \leq T_m^\mathrm{end}  \right \}$, where $T_m^0$ and $T_m^\mathrm{end} $ are the start and end time instants of $\bm Y_m$, respectively. Hence,  $T_1^0 = 1$, $T_M^\mathrm{end} = K$,   $  T_{m+1}^0  = T_m^\mathrm{end} +1 $, and 
\begin{align*}
1 = T_1^0 < \cdots < T_m^0 < T_m^\mathrm{end}  < T_{m+1}^0 <   \cdots < T_M^\mathrm{end} = K.
\end{align*}
Given this partition, $\bm Y = \bigcup_{m=1}^M \bm Y_m$, and $\bm Y_m \cap \bm Y_n =  \emptyset$ for $\quad m\neq n$, and  $\bm Y_m \in \mathbb{R}^{(T_m^\mathrm{end} -  T_m^0+1)n_y}$.
Now, we consider the full posterior  $\p(\bm\theta|\bm Y)$. Bayes' rule gives
\begin{align}
\p(\bm\theta|\bm Y) &= \p(\bm\theta|\bm Y_{1:M})
\propto p(\bm\theta)\prod_{m=1}^{M}
\p(\bm Y_m|\bm\theta, \bm Y_{1:m-1}).
\label{eq:CasEnKI_recursive_update}
\end{align} 
where $\p(\bm\theta|\bm Y_0) = p(\bm \theta)$ with $\bm Y_0 = \emptyset$. This implies that  the posterior can be updated sequentially as
\begin{align}
\p(\bm\theta|\bm Y_{1:m})
\propto
\p(\bm Y_m|\bm\theta,\bm Y_{1:m-1})
\p(\bm\theta|\bm Y_{1:m-1}),
\label{eq:CasEnKI_recursive_posterior}
\end{align}
for $m=1, \ldots, M$. Both~\eqref{eq:CasEnKI_recursive_update}--\eqref{eq:CasEnKI_recursive_posterior}  suggest a cascade of Bayesian updates: at stage $m$,
$\p(\bm\theta|\bm Y_{1:m-1})$, the posterior obtained from the preceding segments $\bm Y_{1:m-1}$, serves as the
prior and is updated using the new subdataset $\bm Y_m$. Further,  we can   apply likelihood tempering to  the incremental
likelihood $\p(\bm Y_m|\bm\theta, \bm Y_{1:m-1})$ in  stage $m$, following~\eqref{eq:EnKI-full-posterior}--\eqref{eq:tempered_posterior} to  introduce
\begin{align}
p_{\ell+1}(\bm{\theta}\,|\,\bm Y_{1:m})
\propto
\left[
\p(\bm Y_m|\bm\theta,\bm Y_{1:m-1})
\right]^{\alpha_{m,\ell}}
p_\ell(\bm{\theta}\,|\,\bm{Y}_{1:m} ),
\label{eq:tempered_posterior_in_CasEnKI}
\end{align}
for $\ell=0,\ldots,L-1$. Then, following similar lines in~\eqref{eq:EnKI_parameter_update}--\eqref{eq:EnKI_iter}, we obtain the following iterative procedure to update   the parameter ensemble $\{ \bm \theta_{m,\ell}^i\}_{i=1}^{N_e}$ that approximates $p_\ell(\bm{\theta}\,|\,\bm{Y}_{1:m} )$:
\begin{subequations}\label{eq:CasEnKI_overall}
\begin{align} \nonumber
\bm\theta_{m,\ell+1}^{i}
&=
\bm\theta_{m,\ell}^{i}
+
\bm C_{m,\ell}^{\bm\theta\bm Y_m} 
\left(
\bm C_{m,\ell}^{\bm Y_m \bm Y_m}
+
\alpha_{m,\ell}^{-1}\bm\Gamma_m
\right)^{-1}\\
& \quad\quad\quad\quad\quad\quad\quad\quad\quad\quad\quad\quad   \cdot
\left(
\bm Y_m
-
\bm Y_{m,\ell}^{i}
\right),
\label{eq:CasEnKI_update} \\ 
\overline{\bm\theta}_{m,\ell}
&=
\SampleMean
\left(
\bm\theta_{m,\ell}^{1\rightarrow N_e}
\right),
\\
\overline{\bm Y}_{m,\ell}
&=
\SampleMean
\left(
\bm Y_{m,\ell}^{1\rightarrow N_e}
\right),
\label{eq:CasEnKI_empirical_means} \\ 
\bm C_{m,\ell}^{\bm\theta\bm\theta}
&=
\SampleCov
\left(
\bm\theta_{m,\ell}^{1\rightarrow N_e}
\right),
\\
\bm C_{m,\ell}^{\bm\theta\bm Y_m}
&=
\SampleCov
\left(
\bm\theta_{m,\ell}^{1\rightarrow N_e},
\bm Y_{m,\ell}^{1\rightarrow N_e}
\right),
\\
\bm C_{m,\ell}^{\bm Y_m\bm Y_m}
&=
\SampleCov
\left(
\bm Y_{m,\ell}^{1\rightarrow N_e}
\right).
\end{align}
\end{subequations}
Here, for $m=1$, the initial parameter ensemble is drawn
from $p(\bm\theta)$, whereas for $m=2,\ldots,M$,  $\{ \bm \theta_{m,0}^i\}_{i=1}^{N_e}$ is initialized by  $\{ \bm \theta_{m-1,L}^i\}_{i=1}^{N_e}$, and 
$\bm Y_{m,\ell}^{i}$ is the predicted output trajectory for
$\bm Y_m$ associated with $\bm\theta_{m,\ell}^{i}$.   
This procedure refines $\{ \bm \theta_{m, \ell+1}^i \}_{i=1}^{N_e}$    by assimilating $\bm Y_m$ incrementally. It runs over $\bm Y_m$ for $m=1, \ldots, M$ in a cascaded way to assimilate the entire dataset, thus called CasEnKI.  The final parameter
estimate and empirical   covariance are $\widehat{\bm\theta}
=
\overline{\bm\theta}_{M,L}$ and $\bm C^{\bm\theta\bm\theta}
=
\bm C_{M,L}^{\bm\theta\bm\theta}$, 
respectively.

Implementing~\eqref{eq:CasEnKI_overall} requires to construct the output trajectory ensemble $\{\bm Y_{m, \ell}^i\}_{i=1}^{N_e}$, which approximately represents $\p(\bm Y_m|\bm Y_{1:m-1},\bm\theta_{m,\ell}^i)$. Note that $\p(\bm Y_m|\bm Y_{1:m-1},\bm\theta)$ satisfies
\begin{align*}
 \p(\bm Y_m|\bm Y_{1:m-1},\bm\theta) 
& = 
\int
\p(\bm Y_m|\bm x_{T_m^0},\bm\theta) \\
&\quad\quad\quad\quad\quad
\cdot \p(\bm x_{T_m^0}|\bm Y_{1:m-1},\bm\theta)
\,d\bm x_{T_m^0},\\
\p(\bm x_{T_m^0}|\bm Y_{1:m-1},\bm\theta)
&= \int \p(\bm x_{T_m^0}|\bm x_{T_{m-1}^{\mathrm{end}}},\bm\theta ) \\
&\quad\quad\quad 
\cdot 
\p(\bm x_{T_{m-1}^{\mathrm{end}}}|\bm Y_{1:m-1},\bm\theta) \, d \bm x_{T_{m-1}^{\mathrm{end}}}.
\end{align*}
This indicates that we must perform 
state filtering conditioned on each  
$\bm\theta_{m,\ell}^{i}$ to construct $\{\bm Y_{m,\ell}^{i}\}_{i=1}^{N_e}$. To avoid repeating state filtering
over the preceding segments $\bm Y_{1:m-1}$ whenever the parameters are updated,
we use the approximation
\begin{align*}
\p(\bm x_{T_m^0}|\bm Y_{1:m-1},\bm\theta_{m,\ell}^{i})
\approx
\p(\bm x_{T_m^0}|\bm Y_{1:m-1},\bm\theta_{m-1,L}^{i}).
\end{align*} 
The  filtering procedure to compute  $\bm Y_m^i$  is as follows. 
First, suppose that the CasEnKI update at stage $m-1$ has yielded the parameter
ensemble $\{\bm\theta_{m-1,L}^{i}\}_{i=1}^{N_e}$ and, for each
$\bm\theta_{m-1,L}^{i}$, an associated   state ensemble
$\left\{\bm x_{T_{m-1}^{\mathrm{end}} | T_{m-1}^{\mathrm{end}} }^{j,i} \right\}_{j=1}^{N_s}$.
We then initialize   stage $m$  by setting
$\bm\theta_{m,0}^{i}=\bm\theta_{m-1,L}^{i}$ for $i=1,\ldots,N_e$  
and running 
\begin{align}
\bm x_{T_m^0 | T_m^0-1}^{j,i}
=
f\left(
\bm x_{T_{m-1}^{\mathrm{end}} | T_{m-1}^{\mathrm{end}}}^{j,i},
\bm\theta_{m,0}^{i}
\right)
+
\bm w_{T_{m-1}^{\mathrm{end}}}^{j,i},
\label{eq:CasEnKI_boundary_state_initialization}
\end{align}
where
$\bm w_{T_{m-1}^{\mathrm{end}}}^{j,i}{\sim}
\mathcal N(\bm0,\bm Q)$. 
Following the initialization, we leverage the EnKF method to conduct state filtering and output prediction at each iteration $\ell$   according to ~\eqref{eq:EnKF_sate_filtering_output_prediction}--\eqref{eq:state_ensemble_prediction} using  $\bm y_k$ for  $k=T_m^0, \ldots, T_m^{\mathrm{end}}$. This generates an ensemble of   output trajectories $\{ \bm y_{k|k-1}^{j,i} \}_{j=1}^{N_s}$ (refer to~\eqref{eq:EnKF_predicted_measurement}), and averaging them gives  $\bm Y_{m, \ell}^i$ as in~\eqref{eq:output_trajectory_average}. 
 The state filtering procedure and the parameter update procedure alternate over
$\ell=0,\ldots,L-1$.

The resulting CasEnKI-based SSM identification algorithm is
summarized in Algorithm~\ref{alg:CasEnKI}. 
For CasEnKI, a major computational cost arises from~\eqref{eq:CasEnKI_update},
which requires computing
$
\left(
\bm C_{m,\ell}^{\bm Y_m\bm Y_m}
+
\alpha_{m,\ell}^{-1}\bm\Gamma_m
\right)^{-1}
$
at an inversion cost of
$\mathcal O\left((T_m^{\mathrm{end}}-T_m^0+1)^3n_y^3\right)$.
For an even partition, each $\bm Y_m$ has length $K/M$, so the
total complexity of~\eqref{eq:CasEnKI_update} across all $M$ segments
is $\mathcal O(K^3n_y^3/M^2)$. This is much lower than 
$\mathcal O(K^3n_y^3)$ in the monolithic EnKI update by a factor of $M^2$.
Hence, CasEnKI reduces the computation by considerable margins, making it strongly scalable to long data sequences.

\begin{algorithm}[t]
\caption{CasEnKI-Based Nonlinear Stochastic SSM Identification}
\label{alg:CasEnKI}
\small
\begin{algorithmic}[1]

\algstage{Initialization}
\State Partition $\bm Y$ into $\{\bm Y_1, \ldots, \bm Y_M\}$  
\State Draw the initial parameter ensemble 
$\{\bm\theta_{1,0}^{i}\}_{i=1}^{N_e}{\sim} p(\bm\theta)$.

\vspace{3pt}
\algstage{CasEnKI-based parameter identification}
\For{$m=1,\ldots,M$}
    \State Set $\ell=0$ and $t_{m,0}=0$.
    \While {$t_{m,\ell}<1$}
    
    \vspace{3pt}
    \algstage{State filtering}
        \For{$i=1,\ldots,N_e$}
            \State Initialize the state ensemble,   run the   EnKF
            using $\bm Y_m$
            \State Construct
            $\bm Y_{m,\ell}^{i}$  
        \EndFor
        
    \vspace{3pt}
    \algstage{Parameter ensemble update}
        \State Update the parameter ensemble using~\eqref{eq:CasEnKI_overall}
        \State Set $t_{m,\ell+1}=t_{m,\ell}+\alpha_{m,\ell}$
        \State Set $\ell\gets\ell+1$
    \EndWhile
    \State Set $L=\ell$
    
    \vspace{3pt}
    \algstage{Preparation for the next stage in cascade}
    \If{$m<M$}
        \State Initialize the next stage by        $\{\bm\theta_{m+1,0}^{i}\}_{i=1}^{N_e}=\{\bm\theta_{m,L}^{i}\}_{i=1}^{N_e}$ 
    \EndIf
\EndFor

\vspace{3pt}
\algstage{Final estimate}
\State Set $\widehat{\bm\theta} = \overline{\bm \theta}_{M,L}$ 
\State \Return $\widehat{\bm\theta}$

\end{algorithmic}
\end{algorithm}

\section{EnKI-Based SSM Identification for LiBs}
\label{sec:BattX}
In this section, we validate the EnKI-based SSM identification approach by applying it to identify the BattX model for LiBs. This case study includes both simulation-based and experimental validation.  For benchmarking purposes, Appendix~\ref{sec:illustrative_example_first_order} compares the proposed approach with the method in~\cite{SCHON:2011:A} using a first-order SSM, as the latter is not capable of identifying the BattX model.

\subsection{Overview of the BattX Model}
\label{sec:BattX_model}
\begin{figure*}[t]
\centering
\includegraphics[width=\textwidth]{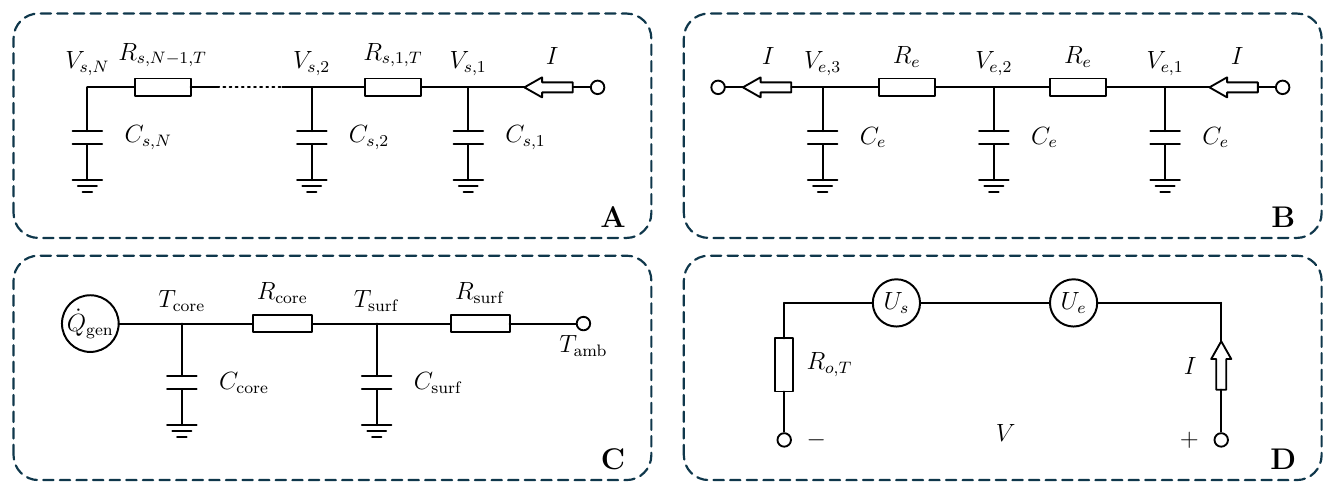}
\caption{The nonlinear BattX model for Li-ion batteries, comprising four coupled sub-circuits: (A) lithium-ion diffusion in the solid electrode, (B) lithium-ion diffusion in the electrolyte, (C) thermal dynamics through heat conduction and convection, and (D) terminal voltage response.}
\label{fig:BattX}
\end{figure*}

The BattX model is an ECM proposed in~\cite{BIJU:2023:AE} to describe the dynamic behavior of a LiB cell operating across a wide range of C-rates. By design, it consists of four coupled sub-circuits, as illustrated in Fig.~\ref{fig:BattX}. Specifically, sub-circuits A and B are RC networks to emulate lithium-ion diffusion in the electrode and electrolyte, respectively; sub-circuit C is a thermal RC network to capture the evolution of the lumped temperature; and sub-circuit D captures the terminal voltage. A more detailed description of the model and its governing equations is available in Appendix~\ref{app:BattX}. Compared with other ECMs, such as the Thevenin model and the nonlinear double-capacitor model~\cite{Tian:2021:TCS}, the BattX model offers higher fidelity by leveraging circuit analogs to capture multiple physical processes within the cell.

For the BattX model, the state, input, and measurement vectors are defined as follows:
\begin{align*}
\bm{x} &= [V_{s,1}\ \cdots \ V_{s,N_{RC}}\ V_{e,1} \ V_{e,2} \ V_{e,3}\ T_\mathrm{core}\ T_\mathrm{surf}]^\top \in \mathbb{R}^{N_{RC}+5} , \\
\bm{u} &= [I\ T_\mathrm{amb}]^\top \in \mathbb{R}^{2}, \\
\bm{y} &= [V\ T_\mathrm{surf}]^\top \in \mathbb{R}^{2},
\end{align*}
where the variables are shown in Fig.~\ref{fig:BattX} and further explained in Appendix~\ref{app:BattX}. The input current $I(t)$ is defined such that $I(t)<0$ during discharge and $I(t)>0$ during charge. Although the BattX model is originally formulated as deterministic, we treat it as a stochastic SSM during identification by introducing a zero-mean white noise process and measurement noise, as in~\eqref{eq:ssm_model}. For tractable identification, we assume that the SoC-OCV relationship is determined from trickle-current charging/discharging tests~\cite{Plett:2015} and that $N_{RC}=5, C_{s,i} = \eta_i C_{s,1}$ and $R_{s,i} = \sigma_i R_{s,1}$, where $\eta_i$ and $\sigma_i$ are known coefficients~\cite{BIJU:2023:AE}. The model then contains 18 unknown parameters to be identified:
\begin{align*}
\bm{\theta} &= \big[ 
C_{s,1}  \quad R_{s,1} \quad C_{e} \quad R_{e} \quad C_{\mathrm{core}} \quad C_{\mathrm{surf}} \quad R_{\mathrm{core}} \quad R_{\mathrm{surf}}\\ 
&\quad\quad\quad\quad \beta_1 \quad \beta_2 \quad \gamma_1 \quad \gamma_2 \quad \gamma_3 \quad  
\kappa_{1} \quad \kappa_{2} \quad c_1 \quad c_2 \quad c_3
\big]^\top,
\end{align*}
where the reader is referred to Appendix~\ref{app:BattX} for the definition of each parameter. 

\subsection{Numerical Simulation}
\label{sec:NS}

For the simulation study, we consider a nominal BattX model and use it to
generate synthetic datasets. The objective is to identify the BattX model
parameters from voltage and surface temperature measurements and compare
the estimates with the nominal parameter values.

The nominal BattX model is taken from~\cite{BIJU:2023:AE}, where it was extracted from experimental data made on a
Samsung INR18650-25R cell with a nominal capacity of $2.5$~Ah, a maximum voltage
of $4.2$~V, and a minimum voltage of $2.5$~V. The nominal parameter values are
listed in Table~\ref{tab:BattX_results}. 

To generate the synthetic datasets, we run the nominal model using diverse current profiles to excite the electrothermal dynamics:
\begin{itemize}
    \item Constant-current discharge profiles at $0.5{\sim} 3$C.
    \item Dynamic profiles including UDDS, LA92, US06, and SC04, scaled to the range of $0{\sim} 5$C.
    \item Pulse-current discharge profiles at $2{\sim} 5$C.
    \item Ambient temperatures ranging from $283$~K to $313$~K.
    \item During the simulations, the process and measurement noises are applied, with the covariances $\bm{Q} = 10^{-8}\bm{I}$ and $\bm{R} = \mathrm{diag}(10^{-4},10^{-3})$, respectively.
\end{itemize}

\begin{figure*}[htbp]
    \centering
    \includegraphics[width=\linewidth]{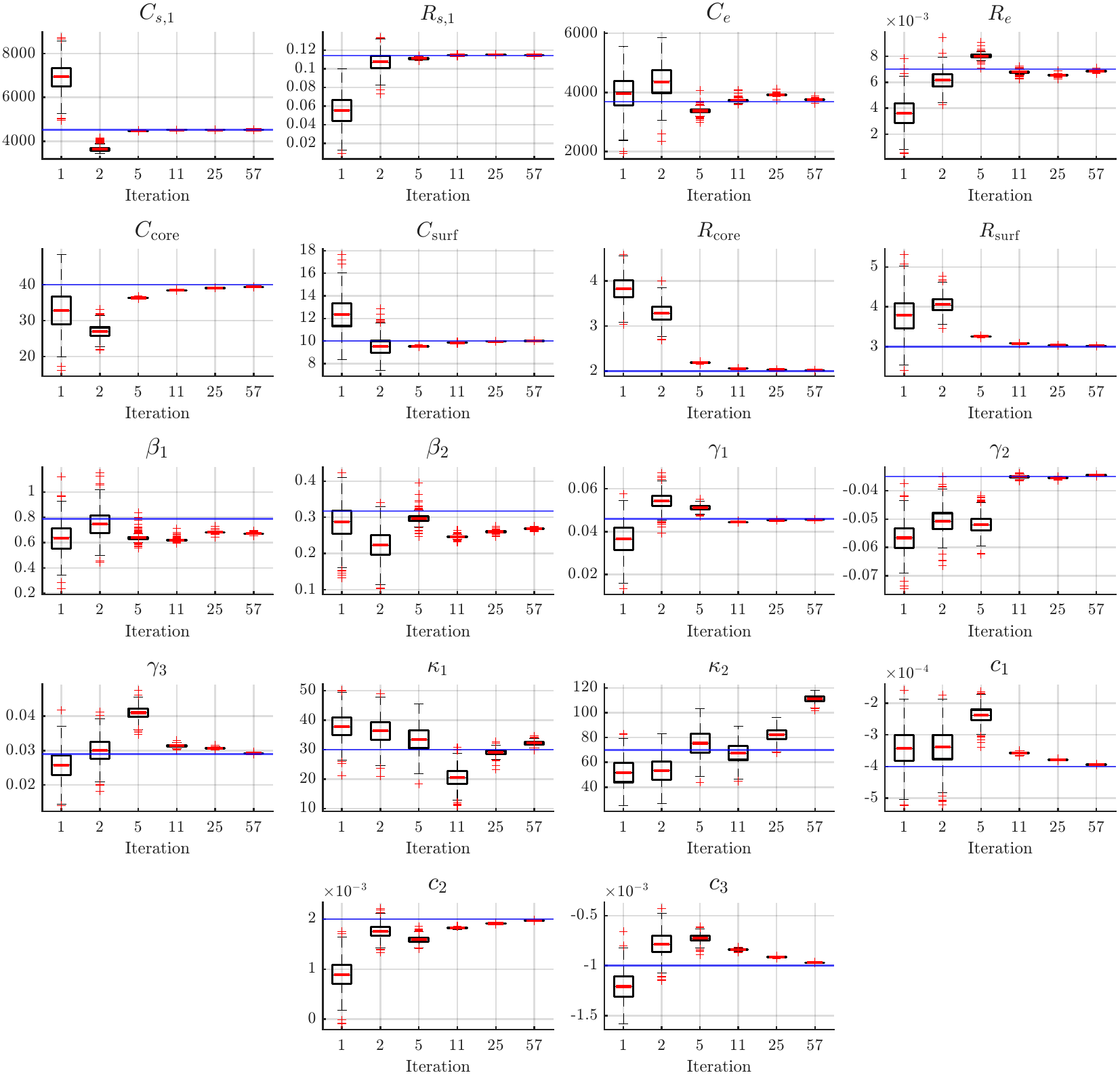}
    \caption{Evolution of the BattX parameter ensembles during CasEnKI iterations. The boxes show the interquartile ranges, whiskers indicate the ensemble spread, red plus markers denote outliers, and horizontal blue lines indicate the true parameter values.}
    \label{fig:boxplot}
\end{figure*}

\begin{figure}[htbp]
    \centering    \includegraphics[width=\columnwidth]{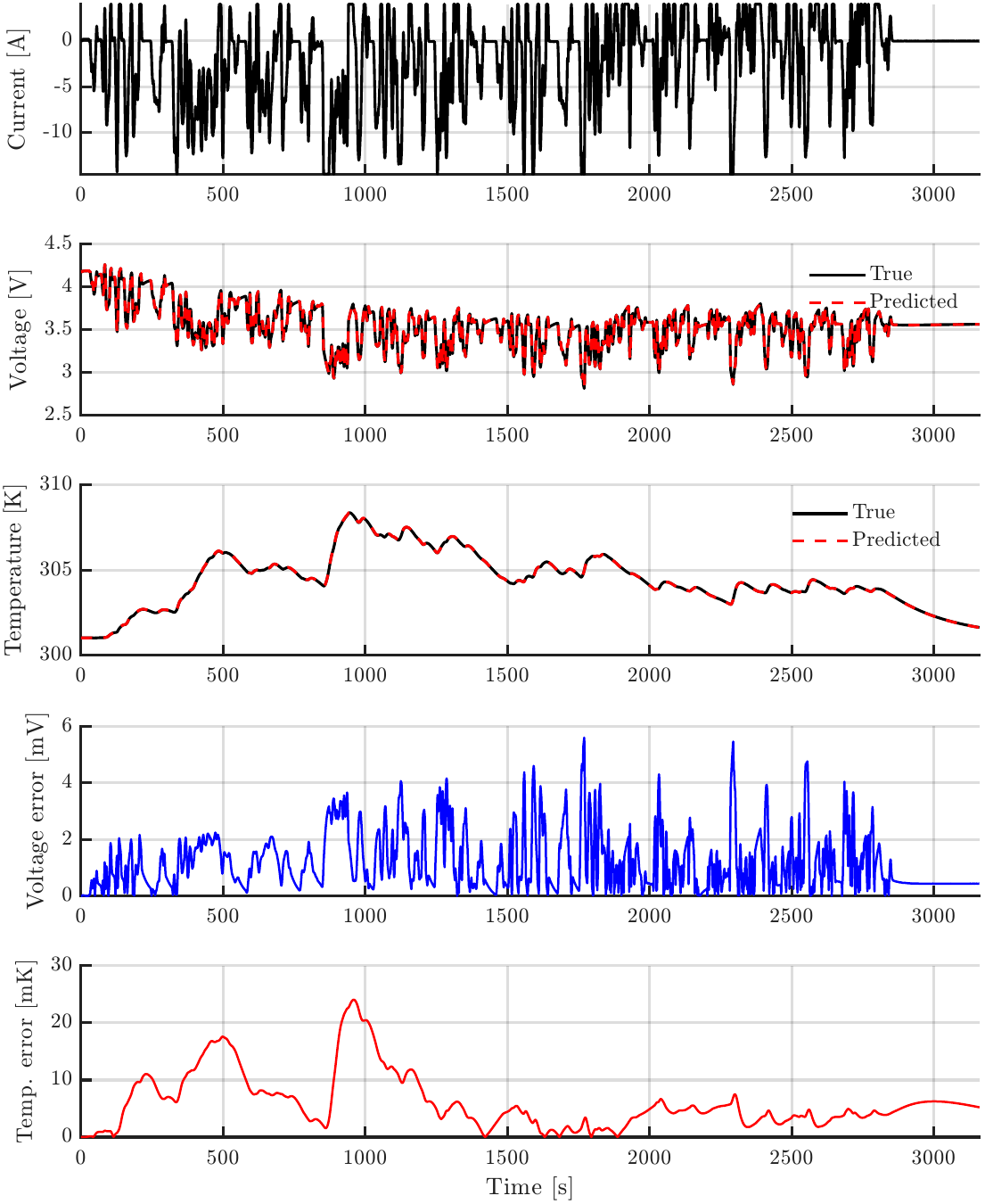}
    \caption{Comparison between the true and identified BattX responses under the LA92 current profile, including current input, terminal voltage, surface temperature, and the corresponding voltage and temperature errors.}
    \label{fig:BattX_Simulation}
\end{figure}

Given the large number of datasets, the EnKI-based identification algorithm becomes computationally unaffordable. We therefore focus on the CasEnKI algorithm. The parameter ensemble size is set to $N_e=500$, with the initial ensemble drawn from a Gaussian prior.
\begin{align*}
\bm\theta_0^{i} {\sim} \mathcal N(\bm\mu_0,\bm C_{0}),
\qquad i=1,\ldots,N_e,
\end{align*} 
where
\begin{align*}
\bm{\mu}_0
&=
\bm{\theta}_{\text{true}}
+
0.4\,\operatorname{diag}\,(\bm{\theta}_{\text{true}})\,\bm{\epsilon},
\qquad
\bm{\epsilon}{\sim}\mathcal{N}(\bm 0,\bm I), \\
\bm C_{0}
&=
\operatorname{diag}\!\left((0.15\,\bm{\theta}_{\text{true}})^2\right).
\end{align*}
Here, $\bm{\theta}_{\text{true}}$ denotes the nominal parameter vector. This initialization generates ensemble members sufficiently different from the nominal values, allowing us to fairly evaluate the CasEnKI algorithm. 
Table~\ref{tab:BattX_results} compares the identified parameter values with their nominal values. For most parameters, the estimation error is below $3\%$. Fig.~\ref{fig:boxplot} further illustrates the evolution of the parameter ensembles during the CasEnKI iterations. Although the initial ensembles deviate substantially from the nominal values, they rapidly contract and converge: their means approach the nominal values within a few iterations while their spreads progressively narrow. Using the identified parameters, we reproduce the BattX model response under the LA92 current profile and compare it with the corresponding measurements in Fig.~\ref{fig:BattX_Simulation}. The identified model closely matches the response of the nominal model, demonstrating the accuracy of the proposed identification approach.

\begin{table}[htbp]
\centering
\ra{1.2}
\caption{Comparison of the true and identified BattX parameter values in simulation.}
 \begin{tabular}{ l | l l l }
\toprule
Parameters & True  & Identified & Error (\%) \\
\midrule
$C_{s,1}$ [F] & 4521 & 4517.582 & 0.08 \\
$R_{s,1}$ [$\Omega$] & 0.1144 & 0.1139 & 0.38\\
$C_e$ [F] & 3691 & 3773.554 & 2.24 \\
$R_e$ [$\Omega$] & 0.007 & 0.0068 & 2.22 \\
$C_{\mathrm{core}}$ [J/K] & 40 & 39.4233 & 1.44 \\
$C_{\mathrm{surf}}$ [J/K] & 10 & 10.0063 & 0.06 \\
$R_{\mathrm{core}}$ [K/W] & 2 & 2.0229 & 1.15 \\
$R_{\mathrm{surf}}$ [K/W] & 3 & 3.0205 & 0.68 \\
$\beta_1$ & 0.789 & 0.6718 & 14.85 \\
$\beta_2$ & 0.317 & 0.2673 & 15.67 \\
$\gamma_1$ & 0.046 & 0.0455 & 1.06 \\
$\gamma_2$ & $-0.035$ & $-0.0345$ & 1.54 \\
$\gamma_3$ & 0.029 & 0.0291 & 0.5 \\
$\kappa_1$ & 30 & 32.0623 & 6.87 \\
$\kappa_2$ & 70 & 108.1826 & 54.55 \\
$c_1$ & $-0.0004$ & $-0.0004$ & 1.43 \\
$c_2$ & $0.002$ & $0.0020$ & 1.37 \\
$c_3$ & $-0.001$ & $-0.0010$ & 2.92 \\
\bottomrule
\end{tabular}
\label{tab:BattX_results}
\end{table}

\begin{figure}[htbp]
\centering
\includegraphics[width=\linewidth]{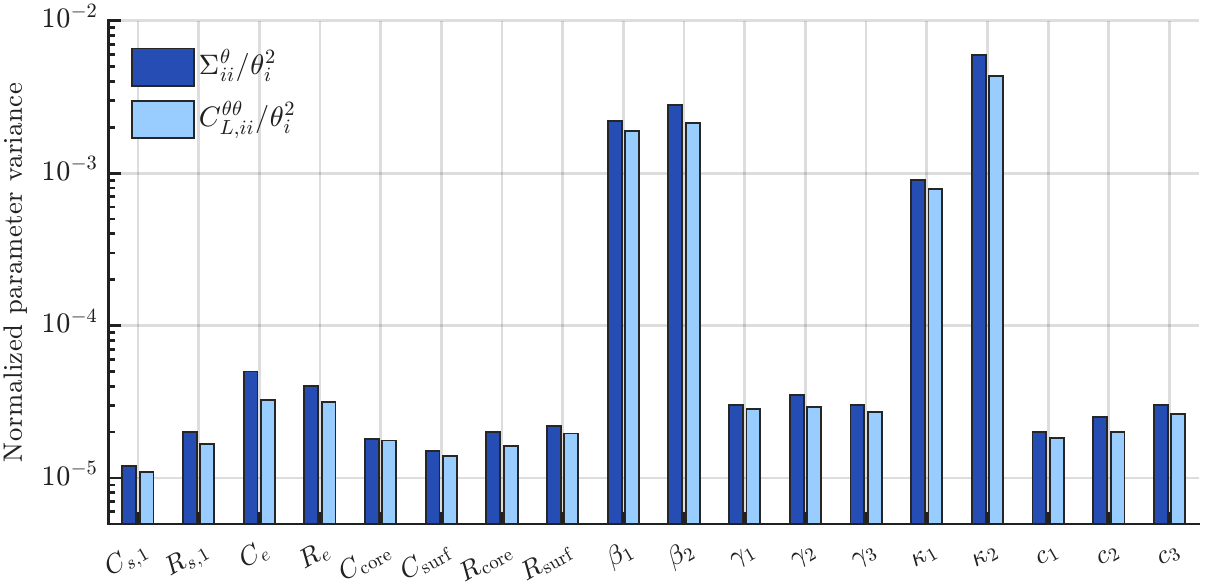}
\caption{Parameter-wise comparison between the diagonal entries of
the sensitivity-based local covariance approximation
$\bm{\Sigma}^{\theta}$ and the final empirical CasEnKI ensemble
covariance $\bm C^{\theta\theta}$.}
\label{fig:sensitivity}
\end{figure}

As shown in Table~\ref{tab:BattX_results} and Fig.~\ref{fig:boxplot}, several parameters, including $\beta_1$, $\beta_2$, $\kappa_1$, and $\kappa_2$, exhibit larger estimation errors than the others. This is because these parameters are less identifiable. To investigate this more, we analyze the parameter identifiability of the BattX model using the framework outlined in Appendix~\ref{app:sensitivity}. The analysis shows that all model parameters are locally identifiable for the given datasets. We further compute the nominal estimation covariance around the nominal parameter vector $\bm\theta$, denoted by $\bm\Sigma^{\theta}$. Fig.~\ref{fig:sensitivity} compares the diagonal entries of $\bm\Sigma^{\theta}$ with those of the parameter covariance matrix $\bm C^{\bm\theta\bm\theta}$ obtained at the final CasEnKI iteration. Here, $\bm\Sigma^{\theta}$ and $\bm C^{\bm\theta\bm\theta}$ represent the sensitivity-based local covariance approximation and the final empirical parameter-ensemble covariance, respectively. As shown in Fig.~\ref{fig:sensitivity}, the two covariance matrices agree closely, further showing the effectiveness of the proposed CasEnKI identification algorithm.

\begin{figure}[htbp]
\centering
\includegraphics[width=0.85\linewidth]{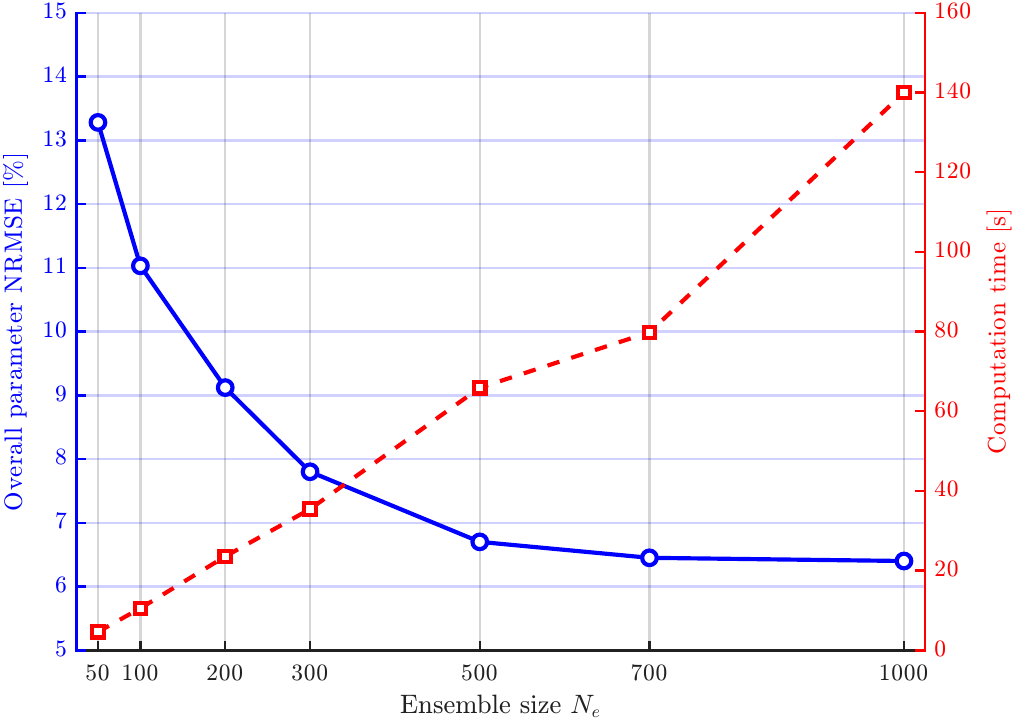}
\caption{Effect of ensemble size on parameter-estimation accuracy and computational time.}
\label{fig:Ensemble_Size_NRMSE_Time}
\end{figure}

We further examine the effect of the parameter ensemble size $N_e$ on estimation accuracy. Fig.~\ref{fig:Ensemble_Size_NRMSE_Time} shows the overall normalized root mean square error (NRMSE) of the parameter estimates and the corresponding computation time for different values of $N_e$. The NRMSE generally decreases as $N_e$ increases, but the gains in accuracy diminish and become marginal beyond $N_e=500$.
Meanwhile, the computation time, measured on a laptop computer equipped with Intel\textsuperscript{\textregistered} Core i7-10700 CPU and 16~GB of RAM, increases substantially. These results indicate a need to balance estimation accuracy and computational cost when selecting $N_e$.

\subsection{Experimental Validation}
\label{sec:Exp}

To further assess its performance, we apply the CasEnKI algorithm to the experimental datasets reported in~\cite{TU:2026:arXiv} to identify the BattX model. The dataset were collected using a Samsung INR18650-25R lithium-ion cell with an NCA cathode and a graphite anode. The cell was tested using an Arbin LBT21084 battery tester inside an ESPEC thermal chamber, and its surface temperature was measured using an attached thermocouple.

Table~\ref{tab:exp_results} lists the search range for each parameter. For the purpose of validation, these ranges are intentionally chosen to be wide, and the initial parameter ensemble is randomly and uniformly generated within them. Table~\ref{tab:exp_results} also reports the parameter values identified by the CasEnKI algorithm. The identified values are consistent with the physical characteristics of the cell and the underlying BattX model.

\begin{table}[htbp]
\centering
\ra{1.2}
\caption{Search ranges and identified BattX parameter values for the experimental validation.}
\begin{tabular}{ l | l l}
\toprule
Parameters & Search range &  Identified \\
\midrule
$C_{s,1}$ [F] & 1000${\sim}$8000 & 6423.2624 \\
$R_{s,1}$ [$\Omega$] & 0${\sim}$1  & 0.6479 \\
$C_e$ [F]  & 1000${\sim}$5000  & 4567.8173 \\
$R_e$ [$\Omega$]  & 0${\sim}$0.1  & 0.0142 \\
$C_{\mathrm{core}}$ [J/K] & 20${\sim}$100 & 59.2864 \\
$C_{\mathrm{surf}}$ [J/K] & 0${\sim}$30  & 16.6883 \\
$R_{\mathrm{core}}$ [K/W] & 0${\sim}$10  & 0.7369 \\
$R_{\mathrm{surf}}$ [K/W] & 0${\sim}$10  & 2.9218 \\
$\beta_1$  & 0${\sim}$200  & 55.9396 \\
$\beta_2$  & 0${\sim}$200  & 133.2875 \\
$\gamma_1$ & 0${\sim}$0.1  & 0.0529 \\
$\gamma_2$ & $-0.1{\sim}0$ & $-0.0311$ \\
$\gamma_3$ & 0${\sim}$0.1  & 0.0266 \\
$\kappa_1$ & 0${\sim}$10000  & 2414.5960 \\
$\kappa_2$ & 0${\sim}$10000  & 5085.8376 \\
$c_1$ & $-0.001{\sim}0$  & $-0.0004$ \\
$c_2$ & 0${\sim}$0.01  & 0.0017 \\
$c_3$ & $-0.01{\sim}0$  & $-0.0012$ \\
\bottomrule
\end{tabular}
\label{tab:exp_results}
\end{table}

\begin{figure*}[t]
    \centering
    \begin{subfigure}[b]{0.48\textwidth}
        \centering
        \includegraphics[width=\textwidth]{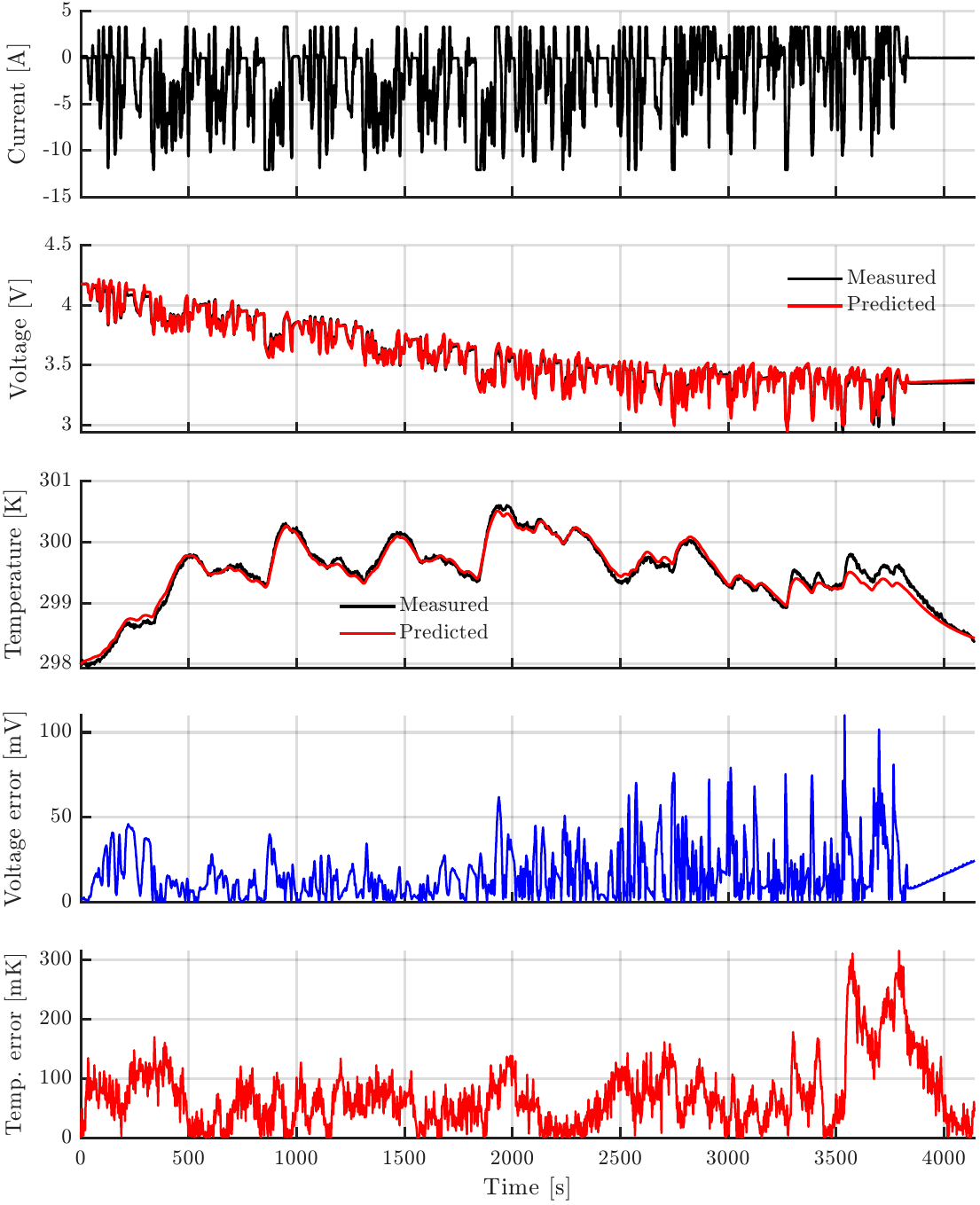}
        \caption{LA92 profile}
        \label{fig:exp_LA92}
    \end{subfigure}\hfill
    \begin{subfigure}[b]{0.48\textwidth}
        \centering
        \includegraphics[width=\textwidth]{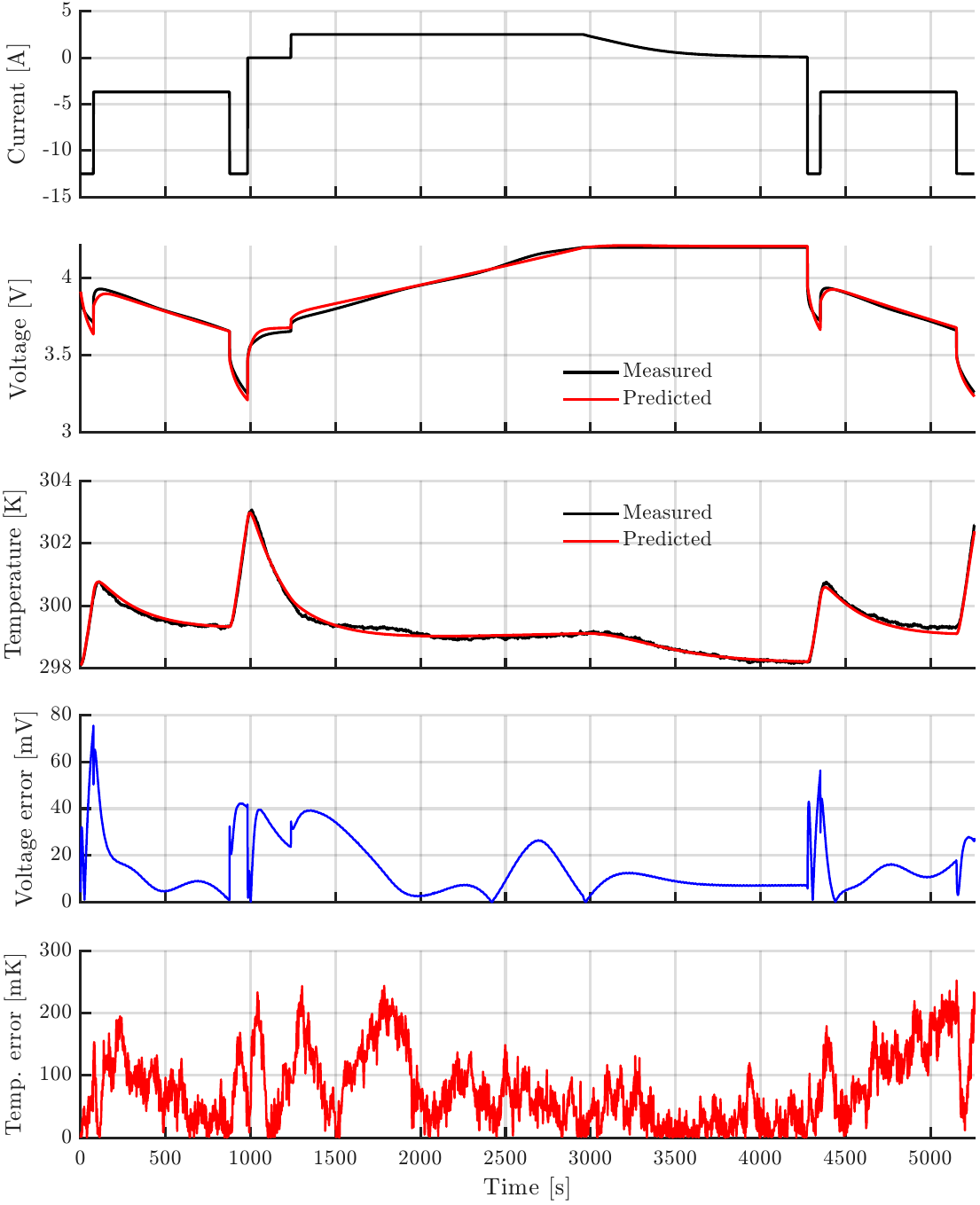}
        \caption{eVTOL profile}
        \label{fig:exp_eVTOL}
    \end{subfigure}
    \caption{Experimental validation of the identified BattX model under two
    operating profiles: (a) LA92 and (b) notional eVTOL. The results compare
    measured and predicted terminal voltage and surface temperature, together
    with the corresponding prediction errors.}
    \label{fig:exp_validation}
\end{figure*}

We further validate the identified BattX model using an experimental dataset collected under the LA92 current profile. Fig.~\ref{fig:exp_LA92} compares the model predictions with the corresponding measurements. Both voltage and temperature predictions show strong agreement with the experimental data, with voltage errors generally below $100\,\mathrm{mV}$ and temperature errors averaging around $200\,\mathrm{mK}$.

We conduct an additional validation using a nominal operating profile representative of an electric vertical take-off and landing (eVTOL) aircraft. This profile is characterized by $5\,\mathrm{C}$ discharge during the takeoff and landing phases. As shown in Fig.~\ref{fig:exp_eVTOL}, the identified BattX model maintains high predictive accuracy, with average voltage and temperature errors of approximately $50\,\mathrm{mV}$ and 
$100\,\mathrm{mK}$, respectively.

\section{Conclusion}
\label{sec:con}

Nonlinear SSMs play a central role in describing dynamic systems across numerous engineering domains. However, identifying their parameters from measurement data has persisted as a challenge for decades. The challenges arise from the combined effects of nonlinear dynamics, latent state variables, and the lack of analytical gradients or the difficult in computing them analytical gradients. The problem becomes even more difficult for high-dimensional SSMs with many unknown parameters. To address these challenges, this paper proposes an EnKI-based identification framework for nonlinear stochastic SSMs. The proposed method maintains an ensemble representation of the posterior distribution of the unknown parameters and iteratively updates the ensemble until convergence. At each iteration, EnKF performs Bayesian state filtering to construct the predicted output trajectory ensemble that is used for the EnKI parameter update. To scale up the method for a large dataset, we further develop a CasEnKI algorithm, which partitions the dataset into smaller subsets and assimilates them sequentially to reduce the computational cost. The resulting framework enables derivative-free parameter estimation, accommodates strong nonlinearity, and offers computational scalability. We validate the framework by identifying the BattX model, a tenth-order ECM with 18 unknown parameters for lithium-ion batteries. Extensive simulation and experimental results demonstrate the accuracy, computational efficiency, and scalability.


\appendices
\counterwithin{equation}{section}  

\section{Sample Mean and Covariance}
\label{app:sample_mean_cov}

Let $\bm x$ be a continuous random vector, and suppose that $N$ samples are drawn from its distribution, denoted by $\bm x^i$, $i=1,\ldots,N$, or collectively by $\bm x^{1\rightarrow N}$. The sample mean and covariance of $\bm x$ are defined as
\begin{align*}
\SampleMean \left(\bm x^{1\rightarrow N}\right)
&=
\frac{1}{N}\sum_{i=1}^{N}\bm x^i,\\
\SampleCov \left(\bm x^{1\rightarrow N}\right)
&=
\frac{1}{N-1}
\sum_{i=1}^{N}
(\bm x^i-\bm m_{\bm x})
(\bm x^i-\bm m_{\bm x})^\top,
\end{align*}
where $\bm m_{\bm x}$ denotes the sample mean of $\bm x$. Likewise, let $\bm y^{1\rightarrow N}$ denote $N$ samples of another random vector $\bm y$. The sample cross-covariance between $\bm x$ and $\bm y$ is defined by
\begin{align*}
&\SampleCov \left(\bm x^{1\rightarrow N},\bm y^{1\rightarrow N}\right)
=\\
&\qquad\qquad\qquad\qquad\qquad
\frac{1}{N-1}
\sum_{i=1}^{N}
(\bm x^i-\bm m_{\bm x})
(\bm y^i-\bm m_{\bm y})^\top,
\end{align*}
where $\bm m_{\bm y}$ is the sample mean of $\bm y$.

\section{An Illustrative Comparison with~\cite{SCHON:2011:A}}
\label{sec:illustrative_example_first_order}
To evaluate the proposed EnKI and CasEnKI SSM identification algorithms, we compare them with the method in~\cite{SCHON:2011:A}. That method combines the EM algorithm with particle smoothing to perform ML-based identification of SSMs and is therefore referred to as particle EM. The comparison focuses on two aspects: estimation accuracy and computational efficiency.

Following~\cite{SCHON:2011:A}, we consider the SSM:
\begin{align*}
x_{k+1}
&= a x_k + b\frac{x_k}{1+x_k^2} + c\cos(1.2k) + w_k,
\\
y_k &= d x_k^2 + v_k,
\end{align*}
where $w_k{\sim}\mathcal{N}(0,q)$ and $v_k{\sim}\mathcal{N}(0,r)$, with $q$ and $r$ known. The unknown parameters to estimate are
\begin{align*}
\bm\theta
=
\begin{bmatrix}
a & b & c & d
\end{bmatrix}^{\top}. 
\end{align*}
The true parameter values are shown in Table~\ref{tab:nonlinear_ssm_results}. We simulate the SSM forward for 1000 time steps to generate a synthetic dataset, which is then used by all the considered algorithms. While EnKI and particle EM both use the entire dataset at every iteration, CasEnKI partitions the dataset into five non-overlapping subdatasets, each with a length of 200. Both EnKI and CasEnKI use a parameter ensemble of 1000 members, and within each, the EnKF procedure has a state ensemble of 100 members. Particle EM employs bootstrap particle filtering and smoothing using 100 particles.

\begin{figure*}[htbp]
\centering
\includegraphics[width=\linewidth]{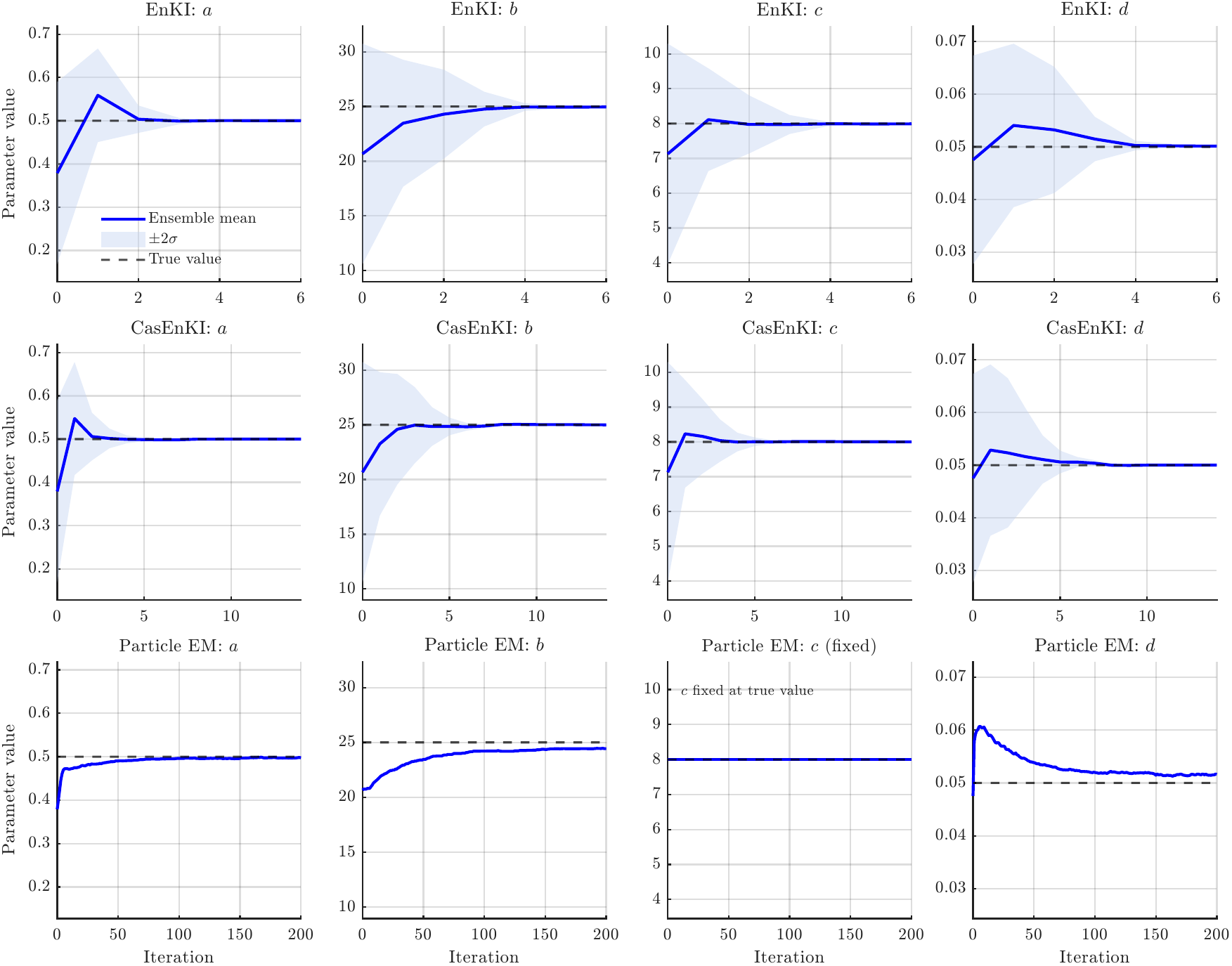}
\caption{Convergence of the parameter estimates for EnKI, CasEnKI,
and particle EM in the nonlinear SSM example. For EnKI and CasEnKI,
the solid curves represent the ensemble means and the shaded regions
indicate $\pm 2$ standard deviations of the parameter ensembles.
The dashed horizontal lines indicate the true parameter values.
For particle EM, the curves show the parameter estimates over the
iterations; the parameter $c$ is fixed at its true value. Each method
is shown using its own iteration axis.}
\label{fig:mean_uncertainty_convergenc}
\end{figure*}

Fig.~\ref{fig:mean_uncertainty_convergenc} compares the parameter estimation results obtained by the three methods. Overall, all methods converge toward the true parameter values as the iterations proceed. However, their convergence behavior differs. EnKI converges in approximately five iterations, while CasEnKI converges after processing all subdatasets in a total of about ten iterations. In contrast, particle EM exhibits much slower convergence, requiring approximately 200 iterations to achieve sufficiently comparable estimation accuracy.

Table~\ref{tab:nonlinear_ssm_results} provides a quantitative comparison of the final estimation errors. EnKI and CasEnKI achieve comparable estimation accuracy, with average relative errors of approximately $0.009\%$ for both methods. Although particle EM is run for substantially more iterations, its errors are considerably larger, at approximately $0.4\%$. These results demonstrate the superior effectiveness of the EnKI-based parameter search over the gradient-based optimization employed by particle EM. One contributing factor is that the parameter ensemble provides richer information about the optimization landscape, leading to more informed parameter updates.

\begin{table*}[htbp]
\centering
\caption{Parameter estimates obtained by EnKI, CasEnKI, and particle EM
for the nonlinear SSM example. Particle EM estimates $a$, $b$, and $d$, while $c$ is fixed at its true value.}
\label{tab:nonlinear_ssm_results}
\renewcommand{\arraystretch}{1.15}
\setlength{\tabcolsep}{6pt}
\small

\begin{tabular}{@{}cc*{3}{cc}@{}}
\toprule
\multirow{2}{*}{$\theta$}
& \multirow{2}{*}{  True Value  }
& \multicolumn{2}{c}{EnKI}
& \multicolumn{2}{c}{CasEnKI}
& \multicolumn{2}{c}{Particle EM} \\
\cmidrule(lr){3-4}
\cmidrule(lr){5-6}
\cmidrule(l){7-8}
&
& Estimate & Error (\%)
& Estimate & Error (\%)
& Estimate & Error (\%) \\
\midrule

$a$ & $0.5$  & $0.500036$  & $0.007$ & $0.50005$   & $0.010$ & $0.505411$ & $1.082$ \\

$b$ & $25$   & $25.0001$   & $0.000$ & $25.0009$   & $0.004$ & $25.0213$  & $0.085$ \\

$c$ & $8$    & $7.99852$   & $0.019$ & $7.99855$   & $0.018$ & $8$ (fixed)        & $0.000$ \\

$d$ & $0.05$ & $0.0500046$ & $0.009$ & $0.0500011$ & $0.002$ & $0.0499745$ & $0.051$ \\

\bottomrule
\end{tabular}
\end{table*}

\begin{figure}[htbp]
\centering
\includegraphics[width=\linewidth]{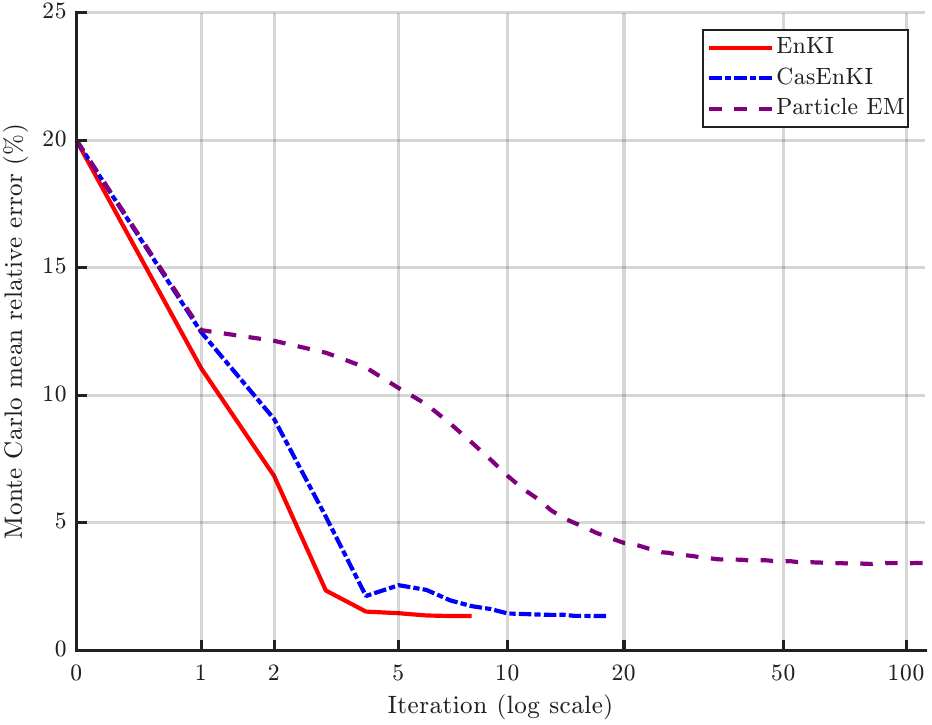}
\caption{Mean relative parameter-estimation error over the common
identified parameters, averaged over 50 Monte Carlo runs, for EnKI, CasEnKI, and particle EM as a function of iteration number. The horizontal axis is shown on a logarithmic scale.}
\label{fig:Comparison_Mean_Error_vs_Normalized_Progress}
\end{figure}

To further evaluate the performance, Fig.~\ref{fig:Comparison_Mean_Error_vs_Normalized_Progress} presents the mean relative errors over 50 Monte Carlo runs, and Table~\ref{tab:overall_comparison} summarizes the corresponding quantitative results together with the average CPU time. The mean relative error for each Monte Carlo run is defined as
\begin{align*}
\mathrm{MRE}
=
\frac{1}{n_\theta}
\sum_{i=1}^{n_\theta}
\frac{
\left|\widehat{\theta}_i-\theta_i\right|
}{
\left|\theta_i\right|
}
\times 100.
\end{align*}
The Monte Carlo results are consistent with the observations above. In terms of computational efficiency, EnKI requires an average CPU time of $0.53946$~s (measured on a laptop computer equipped with an Intel\textsuperscript{\textregistered} Core i7-10700 CPU and 16~GB of RAM), whereas CasEnKI requires only $0.2175$~s, representing a reduction of more than 60\%. As discussed in Section~\ref{sec:CasEnKI}, the computational advantage of CasEnKI becomes even more pronounced for higher-dimensional SSMs and larger datasets. In comparison, particle EM requires an average CPU time of $10.864$~s, which is substantially larger than that of the proposed methods. This high computational cost stems from two factors: particle EM requires substantially more iterations and performs particle smoothing rather than state filtering during each iteration. Consistent with our experimental observations, its computational burden scales poorly with both the model dimension and the dataset size.

\begin{table}[htbp]
\centering
\caption{Monte Carlo comparison of the parameter-estimation accuracy
and computational cost of EnKI, CasEnKI, and particle EM.}
\label{tab:overall_comparison}
\renewcommand{\arraystretch}{1.0}
\setlength{\tabcolsep}{4pt}
\small

\begin{tabular}{lcc}
\toprule
Method
& Final mean relative error (\%)
& CPU time (s) \\
\midrule

EnKI
& $1.3333$
& $0.5395$ \\

CasEnKI
& $1.342$
& $0.2175$ \\

Particle EM
& $3.4223$
& $10.864$ \\

\bottomrule
\end{tabular}
\end{table}

\section{BattX Model: State-Space Formulation}
\label{app:BattX}
As illustrated in Fig.~\ref{fig:BattX}, the BattX model incorporates four coupled sub-circuits as explained below.

Sub-circuit A mimics lithium-ion diffusion in the electrode phase. The capacitors $C_{s,i}$ represent the storage spaces within the electrode where lithium ions intercalate into or disintercalate from the host structure, while the resistors $R_{s,i}$ represent the resistance to lithium-ion transport. Accordingly, charge transfer among $C_{s,i}$ resembles the diffusion of lithium-ions within the electrode. The voltage terms $V_{s,i}$ correspond to lithium-ion concentrations within the electrode. The governing equations of sub-circuit A are given by
\begin{align*}
\dot{V}_{s,1} &= \frac{V_{s,2} - V_{s,1}}{C_{s,1} R_{s,1,T}} + \frac{I(t)}{C_{s,1}}, \\
\dot{V}_{s,i} &= \frac{V_{s,i-1} - V_{s,i}}{C_{s,i} R_{s,i-1,T}} + \frac{V_{s,i+1} - V_{s,i}}{C_{s,i} R_{s,i,T}}, \quad i=2,\dots,N-1, \\
\dot{V}_{s,N} &= \frac{V_{s,N-1} - V_{s,N}}{C_{s,N} R_{s,N-1,T}},
\end{align*}
where $0 \leq V_{s,i} \leq 1$. The SoC then can be expressed as
\begin{align*}
\mathrm{SoC}(t) = \frac{\sum_{i=1}^{N} C_{s,i} V_{s,i}(t)}{\sum_{i=1}^{N} C_{s,i}}.
\end{align*}
The resistances $R_{s,i}$ are temperature-dependent according to
\begin{align*}
R_{s,i,T} = R_{s,i}\,
\exp \left[
\kappa_2
\left(
\frac{1}{T_{\mathrm{core}}}
-
\frac{1}{T_{\mathrm{ref}}}
\right)
\right],
\quad i=1,\dots,N-1,
\end{align*}
where $T_{\mathrm{ref}}=298$~K is the reference temperature and
$\kappa_2$ is the temperature-dependence coefficient.

Sub-circuit~B models lithium-ion transport in the electrolyte, governed by
\begin{align*}
\dot{V}_{e,1} &= \frac{V_{e,2} - V_{e,1}}{C_e R_e} + \frac{I(t)}{C_e}, \\
\dot{V}_{e,2} &= \frac{V_{e,1} - 2V_{e,2} + V_{e,3}}{C_e R_e}, \\
\dot{V}_{e,3} &= \frac{V_{e,2} - V_{e,3}}{C_e R_e} - \frac{I(t)}{C_e}.
\end{align*}

Sub-circuit C captures the cell’s lumped thermal dynamics using the core temperature $T_\mathrm{core}$ and the surface temperature, $T_\mathrm{surf}$. The RC network, composed of the thermal capacitances $C_\mathrm{core}$ and $C_\mathrm{surf}$ and the thermal resistances $R_\mathrm{core}$ and $R_\mathrm{surf}$, represents heat conduction within the cell and convection between the cell surface and the ambient environment. Sub-circuit C is governed by
\begin{align*}
\dot{T}_{\mathrm{core}} &= \frac{\dot{Q}(t)}{C_{\mathrm{core}}} 
+ \frac{T_{\mathrm{surf}} - T_{\mathrm{core}}}{R_{\mathrm{core}} C_{\mathrm{core}}},
\\
\dot{T}_{\mathrm{surf}} &= \frac{T_{\mathrm{amb}} - T_{\mathrm{surf}}}{R_{\mathrm{surf}} C_{\mathrm{surf}}}
- \frac{T_{\mathrm{surf}} - T_{\mathrm{core}}}{R_{\mathrm{core}} C_{\mathrm{surf}}},
\end{align*}
Here, the heat generation rate $\dot{Q}(t)$ is given by
\begin{align*}
\dot{Q}(t) = I(t) \bigl[ V(t) - U_s(\mathrm{SoC}) \bigr] 
+ I(t) T_{\mathrm{core}}(t) \frac{dU_s}{dT_{\mathrm{core}}},
\end{align*}
where $U_s(\cdot)$ is the solid-phase OCV, and
\begin{align*}
\frac{dU_s}{dT_{\mathrm{core}}} = c_1 + c_2 \mathrm{SoC} + c_3 \mathrm{SoC}^2,
\end{align*}
with $c_1$, $c_2$, and $c_3$ being empirical coefficient parameters.

Sub-circuit D represents the cell's terminal voltage:
\begin{align*}
V(t) = U_s(V_{s,1}) 
+ U_e(V_{e,1}, V_{e,3}) 
+ R_{o,T}(\mathrm{SoC}, T_{\mathrm{core}}) I(t),
\end{align*}
where $U_s$ is the electrode's OCV, $U_e$ the potential difference within the electrolyte, and $R_{o,T}$ is the internal resistance. Further, $U_e$ is expressed as
\begin{align*}
U_e(V_{e,1},V_{e,3}) = \beta_1 \ln \left( \frac{V_{e,1} + \beta_2}{V_{e,3} + \beta_2} \right),
\end{align*}
with empirical parameters $\beta_1,\beta_2$.
Moreover, $R_{o,T}$ is both SoC- and temperature-dependent, following
\begin{align*}
R_{o,T} = R_o(\mathrm{SoC}) 
\exp \left[ \kappa_1 \left( \frac{1}{T_{\mathrm{core}}} - \frac{1}{T_{\mathrm{ref}}} \right) \right],
\end{align*}
where $\kappa_1$ is a coefficient and $R_o(\mathrm{SoC})$ is approximated as
\begin{align*}
R_o(\mathrm{SoC}) = \gamma_1 + \gamma_2 \,\mathrm{SoC} + \gamma_3 \,\mathrm{SoC}^2,
\end{align*}
where $\gamma_1,\gamma_2,\gamma_3$ are SoC coefficients.

\section{Identifiability Analysis for SSMs}
\label{app:sensitivity}

Identifiability analysis characterizes the fundamental limits of SSM identification. In this study, we use it to address two key questions: given an SSM and the available datasets, can the model parameters be uniquely determined using the measurement data, and if so, what level of identification accuracy is fundamentally achievable? 
Borrowing from~\cite{VANDOREN:2009:IFAC}, we first present the notion of local identifiability and then show the corresponding analysis.

Although the parameter identification problem is formulated for a stochastic SSM in the paper, identifiability analysis for general stochastic nonlinear systems is difficult. We hence consider the deterministic version for tractable analysis: 
\begin{subequations}
\label{eq:det_ssm}
\begin{align}
\bm{x}_{k+1} &= f(\bm{x}_k,\bm{u}_k,\bm{\theta}), \label{eq:det_ssm_state}\\
\bm{y}_k &= h(\bm{x}_k,\bm{u}_k,\bm{\theta}). \label{eq:det_ssm_output}
\end{align}
\end{subequations}
For convenience, the model is represented by the input/output mapping with clear parameter dependence: 
\begin{align}\label{compact-input-output-model}
\bm Y = \bm H   \left(\bm \theta; \bm U, \bm x_0\right),
\end{align}
where $\bm x_0$ is known, $\bm \theta \in \bm \Theta$, and 
\begin{align*}
\bm Y &=
\begin{bmatrix}
\bm y_1^\top & \cdots & \bm y_K^\top
\end{bmatrix}^{\top}, \\
\bm U &=
\begin{bmatrix}
\bm u_1^\top & \cdots & \bm u_K^\top
\end{bmatrix}^{\top}. 
\end{align*}
The model structure $\bm H   \left(\bm \theta; \bm U, \bm x_0\right)$ is said to be \emph{locally identifiable} at $ \bm{\theta}_m \in \bm \Theta$ for a given $\bm U$ and $\bm x_0$, if for any $\bm{\theta}_1$, $\bm{\theta}_2$ in the neighborhood of $\bm{\theta}_m$, it holds that
\begin{align*}
\bm H   \left(\bm \theta_1; \bm U, \bm x_0\right) = \bm H   \left(\bm \theta_2; \bm U, \bm x_0\right)   \  \Rightarrow  \ \bm{\theta}_1 = \bm{\theta}_2.
\end{align*} 

Given the measurement data $\bm Y$, consider the following problem to achieve parameter estimation:
\begin{align*}
\widehat {\bm \theta} &= \arg\min_{\bm \theta} V(\bm \theta), 
\end{align*}
where 
\begin{align*}
 V(\bm \theta) &= \frac{1}{2} \left[ \bm Y - \bm H   \left(\bm \theta; \bm U, \bm x_0\right) \right]^\top \bm R^{-1} \left[ \bm Y - \bm H   \left(\bm \theta; \bm U, \bm x_0\right) \right] \\ 
& \quad   + \frac{1}{2} ( \bm \theta -  \bm m_0 )^\top \bm P_0^{-1} ( \bm \theta -  \bm m_0 ).
\end{align*}
Here, $\bm R$ is a positive definite matrix for weighting, and the second term of $V(\bm \theta)$ provides regularization. In an MAP setting, $\bm R$ is the covariance of the measurement noise, and $\bm m_0$ and $\bm P_0$ are the prior mean and covariance of $\bm \theta$, respectively. 

The Jacobian and Hessian of $V(\bm \theta)$ are given by
\begin{align*}
{\partial V(\bm \theta) \over  \partial \bm \theta} &= - {\partial \bm H^\top (\bm \theta) \over  \partial \bm \theta} \bm R^{-1} \left[\bm Y - \bm H(\bm \theta) \right] + \bm P_0^{-1} ( \bm \theta -  \bm m_0 ),\\ 
{\partial^2 V(\bm \theta) \over  \partial \bm \theta^2}  &= {\partial \bm H^\top (\bm \theta) \over  \partial \bm \theta} \bm R^{-1}  {\partial \bm H (\bm \theta) \over  \partial \bm \theta} + \bm S + \bm P_0^{-1},
\end{align*}
where $\bm H(\bm \theta)$ is shorthand for $\bm H   \left(\bm \theta; \bm U, \bm x_0\right)$, and $\bm S$ collects the second-order terms in ${\partial^2 V(\bm \theta) \over  \partial \bm \theta^2}$. At $\widehat{\bm \theta}$, ${\partial V (\bm \theta) \over  \partial \bm \theta} = \bm 0$.
Under weak regularization ($\bm P_0^{-1}\approx\bm0$) and neglecting $\bm S$,   $\widehat {\bm \theta}$ will be locally unique. If $\partial \bm H (\bm \theta) \over  \partial \bm \theta$ is of full column rank as the Hessian at $\widehat {\bm \theta}$ is guaranteed to be positive definite. This sufficient condition can be used to check the local identifiability at $\widehat {\bm \theta}$. 

The achievable estimation accuracy is quantifiable by the covariance of $\widehat  {\bm \theta}$, if $\widehat {\bm \theta}$ recovers the true parameters: 
\begin{align*}
\mathrm{Cov}(\widehat {\bm \theta}) =  \bm J^{-1},
\end{align*}
where $\bm J$ is the Fisher information matrix  
\begin{align*}
\bm J = \mathrm{E} \left( \left. \partial^2 V(\bm \theta) \over  \partial \bm \theta^2 \right|_{\widehat {\bm \theta}} \right) . 
\end{align*}
Neglecting $\bm S$ yields the commonly used approximation
\begin{align*}
{\partial^2 V(\bm \theta) \over  \partial \bm \theta^2 } \approx {\partial \bm H^\top (\bm \theta) \over  \partial \bm \theta} \bm R^{-1}  {\partial \bm H (\bm \theta) \over  \partial \bm \theta} +   \bm P_0^{-1}. 
\end{align*}
Note that
\begin{align*}
{\partial \bm H (\bm \theta) \over  \partial \bm \theta} = \begin{bmatrix}
\left({\partial \bm y_1 \over  \partial \bm \theta}\right)^\top &
\cdots &
\left({\partial \bm y_K \over  \partial \bm \theta}\right)^\top
\end{bmatrix}^\top, 
\end{align*}
where $\partial \bm y_k \over \partial \bm \theta $ represents the sensitivity of $\bm y_k$ with respect to $\bm \theta$. For the $i$-th parameter  $\bm \theta(i)$, $\partial \bm y_k \over \partial \bm \theta(i)$ is given by
\begin{align*}
{\partial \bm y_k \over \partial \bm \theta(i)} = {\partial h (\bm x_k, \bm u_k, \bm \theta)\over \partial \bm x_k} {\partial \bm x_k \over \partial \bm \theta(i)} + {\partial h (\bm x_k, \bm u_k, \bm \theta)\over \partial \bm   \theta(i)},
\end{align*}
where  
\begin{align*}
{\partial \bm x_{k+1} \over \partial \bm   \theta(i)} &= {\partial f (\bm x_k, \bm u_k, \bm \theta)\over \partial \bm x_k} {\partial \bm x_k \over \partial \bm \theta(i)} + {\partial f (\bm x_k, \bm u_k, \bm \theta)\over  \partial \bm \theta(i)}. 
\end{align*}
Since the parameters may differ substantially in magnitude, it is more informative to evaluate the normalized estimation covariance
\begin{align*}
\mathrm{Cov}
\!\left(
\bm\Lambda^{-1}\widehat{\bm\theta}
\right)
=
\bm\Lambda^{-1}
\bm J^{-1}
\bm\Lambda^{-1},
\end{align*}
where $\bm\Lambda=\mathrm{diag}(\widehat{\bm\theta})$. This normalized covariance is independent of the parameter magnitudes, thus providing a more meaningful measure of the estimation accuracy.

\balance
\bibliographystyle{IEEEtran}
\bibliography{references}

\end{document}